\documentclass[11pt,preprint,tightenlines,
showpacs,preprintnumbers,amsmath,amssymb,
superscriptaddress,
a4paper,nofootinbib,longbibliography,prc]{revtex4-2}

\usepackage{nicefrac}

\usepackage{natbib}
\usepackage{amssymb,amsbsy,amsmath,amsfonts}
\usepackage{graphicx}
\usepackage{epsf,epsfig,float,latexsym,amsthm,fancyhdr,rotating}
\usepackage{graphics,psfrag,longtable}
\usepackage{slashed}
\usepackage{ulem}
\usepackage[colorlinks,citecolor=blue,linktoc=all,linkcolor=cyan,urlcolor=blue]{hyperref}
\usepackage{upgreek}

\def\hpm{\hphantom{-}}
\def\beq{\begin{equation}}
\def\eeq{\end{equation}}
\def\bea{\begin{eqnarray}}
\def\eea{\end{eqnarray}}
\def\beqa{\begin{equation}\begin{array}{l}}
\def\eeqa{\end{array}\end{equation}}
\def\eqlab#1{\label{eq:#1}}

\def\bv#1{\boldsymbol{#1}}

\def\barr{\left(\begin{array}{c}}
\def\earr{\end{array}\right)}
\def\bmat{\left(\begin{array}{cc}}
\def\emat{\end{array}\right)}
\def\al{\alpha}

\DeclareMathOperator{\tr}{Tr}
\def\3d{3-D}

\def\piEFT/{$\slashed{\pi}$EFT}
\DeclareMathOperator*{\res}{Res}

\usepackage{xcolor}
\usepackage{bm}
\usepackage{slashed}
\usepackage{graphicx}
\allowdisplaybreaks

\AtBeginDocument{%
    \newwrite\bibnotes
    \def\bibnotesext{Notes.bib}
    \immediate\openout\bibnotes=\jobname\bibnotesext
    \immediate\write\bibnotes{@CONTROL{REVTEX42Control}}
    \immediate\write\bibnotes{@CONTROL{%
    apsrev42Control,author="08",editor="1",pages="1",title="0",year="1"}}
     \if@filesw
     \immediate\write\@auxout{\string\citation{apsrev42Control}}%
    \fi
}%

\makeatletter
\g@addto@macro\bfseries{\boldmath}
\makeatother

\begin{document}
\preprint{MITP/21-026}
\preprint{JLAB-THY-21-3417}

\author{Vadim Lensky}
\affiliation{Institut f\"ur Kernphysik,
 Johannes Gutenberg-Universit\"at  Mainz,  D-55128 Mainz, Germany}
\author{Astrid Hiller Blin}
\affiliation{Theory Center, Thomas Jefferson National Accelerator Facility, Newport News, VA 23606, USA}
\author{Vladimir Pascalutsa}
\affiliation{Institut f\"ur Kernphysik,
 Johannes Gutenberg-Universit\"at  Mainz,  D-55128 Mainz, Germany}

\title{Forward doubly-virtual Compton scattering off an unpolarised deuteron in pionless effective field theory}

\begin{abstract}
We calculate the forward unpolarised doubly-virtual Compton scattering (VVCS) off the deuteron
in the framework of pionless effective field theory, up to next-to-next-to-next-to-leading order (N3LO) for the longitudinal and next-to-leading order (NLO) for the transverse amplitude.
The charge elastic form factor of the deuteron, obtained from the residue of the longitudinal VVCS amplitude, is used to extract the value of the single unknown two-nucleon one-photon contact coupling that enters the longitudinal amplitude at N3LO.
We also study the lowest spin-independent generalised polarisabilities of the deuteron.
The calculated unpolarised VVCS amplitude provides a high-precision model-independent input for a future calculation of the two-photon-exchange correction to the Lamb shift of muonic deuterium.
\end{abstract}

\date{\today}

\maketitle

\tableofcontents
\newpage

\section{Introduction}

The deuteron is a fundamental source of information about the emerging nuclear force and, in the absence of free neutron targets, is often used to study the neutron structure. Recent advances in muonic-atom spectroscopy, by the CREMA Collaboration at PSI, led to presently the most precise determination of the charge radii of the proton~\cite{Pohl:2010zza,Antognini:1900ns}, deuteron~\cite{Pohl1:2016xoo}, helium-3 and 4. In case of the deuteron, the isotopic shift measurement also gives an accurate assessment of a subleading nuclear structure contribution --- the so-called two-photon-exchange (TPE) correction. These accurate measurements provide further challenges for the theoretical description of the low-energy nuclear structure with systematically improvable precision. 

While ab initio QCD calculations of these nuclear-structure quantities are still being out of reach, the method of choice for systematic calculations of the nuclear-structure corrections is effective field theories (EFTs) of the strong interaction.  Specifically, we employ the pionless EFT (\piEFT/)~\cite{Kaplan:1996xu,Kaplan:1998tg,Kaplan:1998we,Chen:1999tn,Beane:2000fx,Bedaque:2002mn,Braaten:2004rn,Platter:2009gz}, where
the pions are heavy and hence the nucleon-nucleon ($NN$) interaction is described by contact interactions organized in powers of nucleon three-momentum $P$. This description is limited to momenta well below the pion mass,  $P\ll m_\pi$, which
should be well suited for atomic calculations, where the momenta are of the order of the inverse Bohr radius $\al m_r$ (with $\alpha$ the fine structure constant and $m_r$ the atomic reduced mass). Thus, typical momenta in a light muonic atom, such as muonic hydrogen ($\mu$H) or deuterium ($\mu$D), are below 1 MeV.
The use of \piEFT/ is further motivated by its simplicity. The contact interactions lead to separable $NN$ potential, that enables algebraic resummation, resulting in closed
analytic expressions for the nuclear force. Furthermore, it is strictly renormalisable (in the EFT sense), gauge invariant and hence exactly fulfills low-energy theorems (LETs) such as the Thomson limit.

The \piEFT/ has already been demonstrated to work very well for low-energy properties of light nuclear systems, in particular, deuteron polarisabilities, the electromagnetic form factors of the deuteron, and the Compton scattering (CS) off the deuteron, see, e.g., Refs.~\cite{Chen:1999tn,Griesshammer:2000mi,Beane:2000fi,Chen:2004wv,Ando:2004mm,Ji:2003ia,Chen:2004fg,Chen:2004wwa}. 
Here we shall compute the forward doubly-virtual Compton scattering (VVCS) amplitude, which contains the deuteron structure information on the aforementioned TPE correction. This provides an alternative route to assessing these corrections: the existing calculations employ the nuclear Hamiltonian approach~\cite{Friar:1977cf,Friar:2013rha,Pachucki:2011xr,Pachucki:2015uga,Hernandez:2014pwa,Hernandez:2017mof}, or use the approach based on dispersion relations, either data-driven~\cite{Carlson:2013xea}, or informed by EFT calculations of the deuteron structure functions~\cite{Hernandez:2019zcm,Acharya:2020bxf,Emmons:2020aov}. 

The predictive orders of \piEFT/ for the TPE corrections are exhausted at next-to-next-to-next-to-leading order (N3LO), motivating our choice of calculating at this high order. The counting for the transverse contribution starts one order higher, at the respective N4LO, so the transverse amplitude can be neglected up to N3LO in the calculation of the TPE corrections. This consideration notwithstanding, a calculation of the transverse contribution allows one to study the generalised deuteron polarisabilities, such as the magnetic dipole polarisability and the generalised Baldin sum rule. It also provides input for verifying the smallness of the transverse contribution in the context of the TPE.

We therefore calculate the longitudinal amplitude to N3LO, and the transverse amplitude up to next-to-leading order (NLO) in the so-called $z$-parametrisation scheme~\cite{Phillips:1999hh}. Despite differing only at higher orders, this scheme has shown to converge better than the $\rho$-parametrisation scheme, in which one chooses to fit the effective range $\rho_d$ at NLO, for those observables that are dominated by the long-range properties of the deuteron wave function. This is achieved by fitting the residue of the scattering amplitude at NLO in the $z$-parametrisation, thus ensuring the correct reproduction of the long-distance piece of the deuteron wave function starting from NLO.

A recent next-to-next-to-leading order (NNLO) \piEFT/ result for the longitudinal deuteron structure function~\cite{Emmons:2020aov} (in this connection, see also Refs.~\cite{Friar:2013rha} and~\cite{Hernandez:2019zcm}) achieved a precision of about $5\%$ for the inelastic part of the TPE correction. The present N3LO calculation of the longitudinal deuteron VVCS amplitude is expected to further improve the theoretical uncertainty. With the \piEFT/ expansion parameter $P/m_\pi\simeq \gamma/m_\pi\simeq 1/3$, where $\gamma\simeq 45$~MeV is the deuteron binding momentum, one expects to achieve a relative precision of the order of $(1/3)^4\simeq 1\%$ --- an estimate of the contributions of higher orders in the \piEFT/ expansion, which is the dominant source of uncertainty. This being compatible with other existing results, our work provides an alternative high-precision and model-independent handle on the TPE corrections in $\mu$D. A detailed study of these corrections, making use of the \piEFT/ results for the deuteron VVCS, will be presented elsewhere~\cite{muDpaper}.

The article is organised as follows. Section~\ref{sec:Framework} concerns the details of the \piEFT/ framework for the calculation of the unpolarised deuteron VVCS amplitude. The results of the calculation are presented, with a detailed description of various contributions, and a discussion of the effect of the $NN$ contact terms entering the calculation, in Sec.~\ref{sec:VVCS_results}. Section~\ref{sec:Deuteron_FF_N3LO} considers the deuteron charge form factor and fitting the unknown
N3LO contact term, using the deuteron charge radius. A study of the generalised deuteron polarisabilities is presented in Sec.~\ref{sec:Polarisabilities}, with a summary following in Sec.~\ref{sec:Conclusion}.

\section{Expansions and power-counting for the deuteron VVCS amplitude}
\label{sec:Framework}

\subsection{Unpolarised VVCS amplitudes}

The main subject of study here is the forward doubly-virtual Compton scattering (VVCS) on the deuteron. We will consider only an unpolarised deuteron. In this case, the
general Lorentz decomposition of the VVCS amplitude is the same for a target with any spin, i.e., just as for the nucleon VVCS,  it decomposes into two scalar amplitudes:
\beq 
T^{\mu\nu} (p,q) = \left( -g^{\mu\nu}+\frac{q^{\mu}q^{\nu}}{q^2}\right)
T_1(\nu, Q^2) +\frac{1}{M_d^2} \left(p^{\mu}-\frac{p\cdot
q}{q^2}\,q^{\mu}\right) \left(p^{\nu}-\frac{p\cdot
q}{q^2}\, q^{\nu} \right) T_2(\nu, Q^2)\,,
\label{eq:VVCS_amplitude}
\eeq
where $q$ and $p$ are the photon and deuteron four-momenta, $M_d$ is the deuteron mass, $\nu=p\cdot q/M_d$ is the photon energy in the deuteron rest frame, and $Q^2=-q^2$ is the photon virtuality. For later use, it is convenient to introduce the longitudinal and transverse amplitudes,
\begin{equation}
    f_L(\nu,Q^2) = -T_1(\nu,Q^2)+\left(1+\frac{\nu^2}{Q^2}\right)T_2(\nu,Q^2)\,, \qquad f_T(\nu,Q^2) = T_1(\nu,Q^2)\,.
\end{equation}
Their interpretation becomes evident by contracting the VVCS tensor with the initial and final photon polarisation vectors (assuming $\epsilon\cdot q=\epsilon'\cdot q =0$):
\bea
T_{fi} & \equiv & \epsilon^{'*}_\mu \epsilon^{\textcolor{white}{'*}}_\nu\, T^{\mu\nu} = -T_1(\nu, Q^2)\, \epsilon\cdot \epsilon^{\,\prime*}
+ T_2(\nu, Q^2)\, \frac{p\cdot \epsilon\, p\cdot \epsilon^{\,\prime*}}{M_d^2} 
\eea
and observing that for the purely longitudinal and transverse photons the amplitude enters in, respectively, the $f_L$ and $f_T$ combination.
We perform the calculation of the VVCS amplitude in the deuteron rest frame, and it is convenient to rewrite it in terms of the time and space components of the photon polarisation vectors in that frame, $\epsilon=(\epsilon_0,\bv{\epsilon})$, which results in
\begin{equation}
    T_{fi} = \varepsilon_0\,\varepsilon_0^{\,\prime*}\,f_{L}(\nu,Q^2) +(\bv{\varepsilon} \cdot \bv{\varepsilon}^{\,\, \prime *}) \,f_{T}(\nu,Q^2)\,,
\label{eq:VVCS_amplitude_LandT}
\end{equation}
where $\varepsilon_0$ and $\bv{\varepsilon}$ are the following combinations of $\epsilon_0$ and $\bv{\epsilon}$ (and analogously for $\varepsilon_0'$ and $\bv{\varepsilon}'$):
\begin{align}
\varepsilon_0&=\left[\epsilon_0-\frac{\nu}{\left|\bv{q}\right|}\, (\bv{\epsilon}\cdot\bv{\hat q})\right]\frac{\left|\bv{q}\right|}{Q}\,,
&\bv{\varepsilon} = \bv{\epsilon}-\bv{\hat q}\,(\bv{\epsilon}\cdot\bv{\hat q})\,,\eqlab{modified_polarization_vectors}
\end{align}
with $\bv{q}$ and $\bv{\hat{q}}=\bv{q}/|\bv{q}|$ the photon three-momentum in the deuteron rest frame and its unit vector. These definitions ensure that $\varepsilon_0$ and $\bv{\varepsilon}$ turn to zero if $\epsilon$ is substituted by $q$. Accordingly, any other deuteron-spin independent structures have to vanish in the sum of a gauge invariant subset of Feynman graphs. This represents an important non-trivial consistency check of our calculation.

The rest of generalities is also very much analogous to the nucleon case (see, e.g., Refs.~\cite{Drechsel:2002ar,Hagelstein:2015egb}). For example, the VVCS amplitudes are split into the pole (or elastic) and non-pole (or inelastic) parts. The former have a pole at $\nu=\pm Q^2/(2M_d)$ and correspond to the VVCS process going through the deuteron in the intermediate state; they are parametrised by the deuteron elastic electromagnetic form factors. The latter admits an expansion in powers of the energies and momenta, and are parametrised by the deuteron (generalised) polarisabilities, as shown below.
In the rest of this section, we briefly recap the essential details of the \piEFT/ expansion, the power-counting, and the Lagrangian needed to
compute the VVCS amplitudes.

\subsection{EFT expansion and counting}
\label{sec:counting}
To set up the counting for the deuteron VVCS, we recall that the \piEFT/ expansion is organised in powers of the ratio $P/m_\pi$, where $P$ is the typical momentum scale in the system. Typical energies are counted as $E\propto P^2$: the two-nucleon system is considered as predominantly non-relativistic, and relativistic corrections are taken into account perturbatively. Correspondingly, a nucleon propagator counts as $1/E\propto P^{-2}$, whereas a loop gives a factor of $P^5$ corresponding to an integration over $d^4P=\mathrm{d}E\, \mathrm{d}^3P$. To assign a particular order to a Feynman graph, one counts powers of momenta coming from the interaction vertices, propagators, and loops, assuming that all momenta are of the typical size $\sim P$ (and all energies of the size $\sim P^2$). 

To further arrive at the counting for the VVCS amplitude, we turn to the non-pole part of the amplitude, and first consider its low-energy and low-momenta expansion. The leading terms of that expansion of the non-pole pieces of $f_L(\nu,Q^2)$ and $f_T(\nu,Q^2)$ are given by the deuteron electric and magnetic dipole polarisabilities $\alpha_{E1}$ and $\beta_{M1}$ as~\cite{Drechsel:2002ar}
\begin{align}
    f_L(\nu,Q^2) &= 4\pi \alpha_{E1}Q^2 + \dots\,,\label{eq:fLLEX}\\
    f_T(\nu,Q^2) &=-\frac{e^2}{M_d}+4\pi \beta_{M1}Q^2 + 4\pi(\alpha_{E1}+\beta_{M1})\nu^2+\dots\,,
\label{eq:fTLEX}
\end{align}
where the dots denote terms at least quartic in $\nu$ and $Q$, and $e$ is the proton charge; the first term in the expansion of $f_T(\nu,Q^2)$ --- the Thomson term --- corresponds to the point-like deuteron.
We neglect other terms generated by the expansion of the non-pole pieces of the Born contributions, such as, e.g., the term $e^2 R_C^2 Q^2/(6M)$, where $R_C$ and $M=(M_n+M_p)/2$ are the deuteron charge radius and the average nucleon mass, respectively, in the expansion of $f_T$~\cite{Gorchtein:2015eoa}; see also, e.g., the related discussion in Ref.~\cite{Birse:2012eb} for the case of a proton. One can show that such terms are all demoted to higher orders in the \piEFT/ counting than those we consider; in any case, they give numerically negligible contributions to the VVCS amplitude. 

Note that we count the photon energy $\nu=O(P^2)$ and momentum $|\bv{q}|=O(P)$. This counting would be questionable if not obviously unsuitable for real photons, however, the case of VVCS implies $Q^2=\bv{q}^2-\nu^2\ge 0$, which is consistent with the assignment we use. Furthermore, this counting is well-suited for an evaluation of the generalised deuteron polarisabilities, where one takes the limit of $\nu\to 0$ keeping $Q^2$ finite; the generalised polarisabilities are obtained as $Q^2$-dependent coefficients of the expansion of the VVCS amplitudes in powers of $\nu^2$. Finally, this counting is compatible with the region of $(\nu, Q^2)$ relevant to the evaluation of the TPE correction in $\mu$D, in other words, the typical energy transfer is considerably less than the momentum transfer. This is evident in the elastic TPE contributions where $\nu = \pm Q^2/(2M_d)$, and it is also true for the inelastic part of the TPE correction (as another reflection of the predominantly non-relativistic character of the deuteron).

Looking at the leading terms in the \piEFT/
expansion of $\alpha_{E1}$ and $\beta_{M1}$~\cite{Chen:1998vi,Phillips:1999hh,Ji:2003ia}, one can see that they are, respectively, $O(P^{-4})$ and $O(P^{-2})$:
\begin{align}
\alpha_{E1} &= \hpm\frac{\alpha M}{32\pi \gamma^4}+\dots\,,\\
\beta_{M1}  &= -\frac{\alpha}{32M\gamma^2}\left[
1 - \frac{16}{3}\mu_1^2 + \frac{32}{3}\mu_1^2\frac{\gamma}{\gamma_s-\gamma}
\right]+\dots\,,
\end{align}
where  $\mu_1$ is the nucleon isovector magnetic moment (in nucleon magneton units), and $\gamma_s\equiv a_s^{-1}=O(P)$ is the inverse proton-neutron singlet scattering length. 

Since $Q^2=\bv{q}^2-\nu^2=O(P^2)$ and $\nu^2=O(P^4)$, we can see, respectively, from Eqs.~\eqref{eq:fLLEX} and~\eqref{eq:fTLEX} that $f_L(\nu,Q^2)$ starts at $O(P^{-2})$, and $f_T(\nu,Q^2)$ starts two orders higher at $O(P^0)$ [to be precise, all terms shown in Eq.~\eqref{eq:fTLEX} are $O(P^0)$, except the last term $\propto \beta_{M1}\nu^2$, which is $O(P^2)$]. This derivation applies to the non-pole parts of the amplitudes. However, it is straightforward to deduce that the same counting holds also for the respective pole parts, which is evident from the expressions for the residues of $f_L(\nu,Q^2)$ and $f_T(\nu,Q^2)$ considered in Sec.~\ref{sec:Deuteron_FF_N3LO}.

Furthermore, in the TPE correction to the deuterium Lamb shift, one can notice that $f_T$ is weighted with another small factor of $O(P^2)$ relative to $f_L$ in the integral for the TPE correction~\cite{Acharya:2020bxf,Carlson:2013xea}. In other words, $f_T$ starts to contribute to the TPE correction only at N4LO relative to the leading longitudinal contribution.

To identify the highest order in the \piEFT/ expansion where one can still hope to obtain a predictive result for the TPE correction, one can notice that the contribution of the polarisabilities of individual nucleons to $f_L$ $\propto \alpha_{E1,N} Q^2$ arises at N4LO in the \piEFT/ counting. This term goes as $Q^2$ at large $Q$ and thus leads to a divergent contribution to the TPE correction, which demands an unknown two-nucleon one-lepton contact term at this order to regularise the divergence. Such a contact term would be fitted to the TPE correction (or another two-nucleon one-lepton observable --- if such data were available). Consequently, the predictive power of \piEFT/ for the TPE correction is lost beyond N3LO. In accordance to that, we concentrate on the longitudinal amplitude and calculate $f_L(\nu,Q^2)$ up to N3LO, or $O(P)$. Taking the \piEFT/ expansion parameter as $\gamma/m_\pi\sim 1/3$, this leads to a na\"ively expected relative uncertainty of the calculation of $\sim\left(\gamma/m_\pi\right)^4\simeq 1\%$. Note that at N3LO there also appears a two-nucleon one-photon coupling which requires input from inelastic processes. We determine its value from the deuteron charge radius in Sec.~\ref{sec:Deuteron_FF_N3LO}, preserving the predictability of the TPE correction results at this order. One has to mention that the extraction of the charge radius from experimental data uses the theoretical prediction for the TPE correction as one of the inputs~\cite{Pohl1:2016xoo}, however, the effect of the considered coupling on the extracted value of the charge radius is rather small, and the undesirable correlation is insignificant. It can be completely avoided if one uses the hydrogen-deuterium isotope shift to extract the charge radius: even though the TPE correction also contributes to the isotope shift, its change due to the coupling in question is far below the current level of precision (at any reasonable value of the coupling constant). This issue is investigated in detail in our subsequent publication~\cite{muDpaper}.

This consideration indicates that the transverse amplitude starts to enter the TPE correction one order higher, and only the longitudinal amplitude is needed in case one is solely aiming at an N3LO calculation of the TPE correction. However, we in addition calculate $f_T(\nu,Q^2)$ up to $O(P)$, or the respective NLO, in order to precisely quantify the smallness of its contribution to the TPE correction, and also to investigate the (generalised) deuteron magnetic polarisability $\beta_{M1}(Q^2)$ and the deuteron generalised Baldin sum rule (as well as its fourth-order analog), associated with the transverse amplitude.

\begin{figure}[htb]
    \centering
    \includegraphics[width=0.85\textwidth]{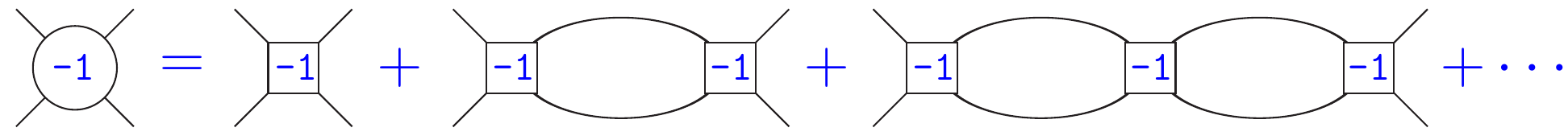}
    \caption{The LO $NN$ $T$-matrix for the spin-triplet channel. Here and below, big disc (square) vertices denote insertions of a $NN$ $T$-matrix ($NN$ potential), with the number inside a vertex showing the order of the vertex in the \piEFT/ counting. The LO $T$-matrix in the spin-singlet ${}^1S_0$ channel is obtained analogously.}
    \label{fig:MLO}
\end{figure}

To conclude the discussion of the \piEFT/ counting and expansion, we consider the $NN$ $T$-matrix in the spin-triplet (deuteron) channel, given by
\begin{align}
    T(k) = -\frac{4\pi}{M}\frac{1}{-\gamma-ik+\frac{1}{2}\rho_d(k^2+\gamma^2)+w_2(k^2+\gamma^2)^2+\dots}\,,
\end{align}
where $k$ is the $NN$ relative momentum, $\rho_d$ and $w_2$ are the deuteron effective range and shape parameter, and terms of higher orders in $(k^2+\gamma^2)$ are not shown explicitly.
Ref.~\cite{Phillips:1999hh} notes that, while the more conventional $\rho$-parametrisation reproduces the deuteron effective range $\rho_d$ at NLO in the \piEFT/ expansion:
\begin{align}
     T(k) &= -\frac{4\pi}{M}\Bigl[
    \underbrace{\frac{1}{-\gamma-ik}}_\text{LO}
    +
    \underbrace{\frac{\gamma\rho_d}{-\gamma-ik}+\frac{\gamma\rho_d}{2\gamma}\, }_{\text{NLO}}+\dots\Bigr]\,,
\end{align}
one can use an alternative scheme, the $z$-parametrisation, choosing instead to reproduce the residue of $T(k)$ at the deuteron pole $k=i\gamma$ at NLO:
\begin{align}
    T(k) & = -\frac{4\pi}{M}\Bigl[
    \underbrace{\frac{1}{-\gamma-ik}}_\text{LO}
    +
    \underbrace{\frac{Z-1}{-\gamma-ik}
    +\frac{Z-1}{2\gamma}}_{\text{NLO}}+\dots\Bigr]\,,
\end{align}
with the residue $Z$ given by
\begin{align}
    Z=\frac{1}{1-\gamma\rho_d}=1+\gamma\rho_d+\left(\gamma\rho_d\right)^2 +\dots\,.
\end{align}
Each scheme introduces a new $O(P)$ small parameter: $\gamma \rho_d$, or $Z-1$.
The leading-order (LO) $O(P^{-1})$ result  is the same in both schemes.
In fact, they only start to differ at NNLO, i.e., $O(P)$. 
Furthermore, both schemes ensure that the deuteron pole is located at $k=i\gamma$ at all orders in the expansion, as obtained by re-summation of an infinite chain of diagrams shown in Fig.~\ref{fig:MLO}.

 The residue $Z$
is connected to $A_S$, the asymptotic normalisation factor of the deuteron $S$-wave, 
 \begin{align}
 \psi(r)\xrightarrow[r\to\infty]{} \frac{A_S}{\sqrt{4\pi}}\frac{e^{-\gamma r}}{r}=\sqrt{\frac{\gamma Z}{2\pi}}\frac{e^{-\gamma r}}{r}\,.
 \end{align}
As argued in Ref.~\cite{Phillips:1999hh}, the $z$-parametrisation is better suited for quantities that receive mostly long-range contributions and are hence sensitive to the correct description of the long-range tail of the deuteron wave function, one example of such a long-range quantity being the deuteron electric polarisability $\alpha_{E1}$. We too adopt the $z$-parametrisation for our calculation.

\subsection{Lagrangian and coupling constants}
The Lagrangian needed for our calculation is constructed along the usual lines formulated in, e.g, Refs.~\cite{Chen:1999tn,Kaplan:1996xu,Kaplan:1998we,Kaplan:1998tg,Chen:1999bg,Rupak:1999rk,Phillips:1999hh}, performing a non-relativistic expansion in the one-nucleon sector and writing out the relevant two-nucleon interactions. The relativistic corrections in both the single-nucleon and  two-nucleon sector count as $O(P^2/M^2)= m_\pi^2/M^2\,O(P^2/m_\pi^2)$ and are therefore more suppressed numerically than suggested by counting powers of momenta. Even when generously counting $m_\pi/M\sim P$, as suggested in, e.g., Ref.~\cite{Rupak:1999rk},  the relativistic corrections start
to appear at N4LO. We therefore neglect them and define $\gamma=\sqrt{-M E_d}$, where $E_d$ is the deuteron energy relative to the proton and neutron at rest. 

We also neglect the isospin violation due to both the proton-neutron mass difference and the isospin-violating terms in the $NN$ interaction, with the caveat that the $NN$ interactions in the singlet channel are fitted to the empirical singlet $pn$ scattering length and effective range and therefore include some isospin-violating effects in that channel.

 The one-nucleon Lagrangian needed for the present VVCS calculation reads
\begin{align}
\mathcal{L}^{N}&= N^\dagger\left[iD_0+\frac{\bv{D}^2}{2M}\right]N
+ \frac{e}{2M} N^\dagger \hat\mu\left(\bv\sigma\cdot\bv B \right)N + \frac{e}{6} N^\dagger\hat{r}_E^2 N\,\left(\bv{\nabla}\cdot\bv{E}\right)\,.
\label{eq:LagN}
\end{align}
The gauge derivatives are defined as
\begin{align}
D_0 N = \left(\partial_0 + i e \hat{Q} A_0\right) N\,,\quad \bv{D} N = \left(\bv{\nabla}-ie\hat{Q}\bv{A}\right)N
\end{align}
with the electromagnetic potential $(A_0,\bv{A})$ and the nucleon charge operator
\begin{align}
\hat{Q} =\frac{1}{2}(1+\tau_3)\,.
\end{align}
The electromagnetic field components are 
\begin{align}
    \bv{B} = \bv{\nabla}\times\bv{A}\,,\quad \bv{E}=-\bv{\nabla} A_0 -\partial_0\bv{A}\,,
\end{align}
and the nucleon magnetic moment and charge radius operators are defined by
\begin{align}
\hat{\mu} = \mu_0+\mu_1\,\tau_3 \,, \qquad \hat{r}^2_E= r_0^2 + r_1^2\,\tau_3\,
\end{align}
where $\mu_\kappa=\nicefrac{1}{2}[\mu_p+(-1)^\kappa \mu_n]$ and $r^2_\kappa=\nicefrac{1}{2}\left[r_p^2+\nicefrac{3}{4}M_p^{-2}+(-1)^\kappa\, r_n^2\right]$ are the nucleon
isoscalar $(\kappa=0)$ and isovector $(\kappa=1)$ magnetic moments (in nuclear magneton units) and charge radii squared, respectively.
The proton charge radius squared appearing here acquires the Darwin-Foldy correction $\nicefrac{3}{4}M_p^{-2}$, see, e.g., Ref.~\cite{Hill:2012rh} for a derivation.

Note that different parts of the minimal charge coupling are of different orders in the \piEFT/ counting: the longitudinal coupling is $\propto A_0=O(P^0)$, whereas the transverse coupling is $\propto \bv{\nabla}\cdot\bv{A}=O(P)$, and the seagull term is $\propto \bv{A}^2=O(P^0)$. Furthermore, while $\bv{B}=O(P)$, $\bv{E}$ contains two parts that are of different orders, $\bv{\nabla}A_0=O(P)$ and $\partial_0\bv{A}=O(P^2)$. The gauge invariance thus mixes different orders in the \piEFT/ counting. This is consistent with the transverse amplitude being suppressed in relation to the longitudinal one.

The Lagrangian describing $NN$ interactions in the triplet $S$-wave up to N3LO and in the singlet $S$-wave up to NLO is given by
\begin{align}
\mathcal{L}^{NN}_S&=-C_0\ N^\dagger \mathcal{P}_i N_c \ N^\dagger_c \mathcal{P}_i N
-C_0^s\ N^\dagger \mathcal{T}_a N_c \ N^\dagger_c \mathcal{T}_a N
\nonumber \\
&+
 \frac{1}{2}C_2\left[N^\dagger \mathcal{P}_i N_c\ N^\dagger_c \mathcal{O}^{(2)}_i N+ \mathrm{H.c.} 
\right]
+
 \frac{1}{2}C_2^s\left[N^\dagger \mathcal{T}_a N_c\ N^\dagger_c \mathcal{O}^{(2,s)}_a N+ \mathrm{H.c.} 
\right]
\nonumber\\
&-C_4
N^\dagger \mathcal{O}^{(2)}_i N_c
\ N^\dagger_c \mathcal{O}^{(2)}_i N
-\frac{1}{2}\tilde{C}_4\left[N^\dagger \mathcal{P}_i N_c\ N_c^\dagger \mathcal{O}^{(4)}_i N + \mathrm{H.c.}\right]
\nonumber \\\
&+\frac{1}{2}C_6\left[N^\dagger \mathcal{O}^{(2)}_i N_c\ N_c^\dagger \mathcal{O}^{(4)}_i N + \mathrm{H.c.}
\right] \,.
\label{eq:LagNN}
\end{align}
Here, we defined the charge-conjugated nucleon field as
\begin{align}
    N_c = \tau_2\,\sigma_2 \left(N^\dagger\right)^T\,;
\end{align}
note that
\begin{align}
\bv{D}N_c = \left(\bv{\nabla}+i e\hat{Q}_c \bv{A}\right)N_c,
\end{align}
with
\begin{align}
    \hat{Q_c} = \tau_2\, Q\,\tau_2 = \frac{1}{2}(1-\tau_3)\,.
\end{align}
The spin-triplet-isospin-singlet and spin-singlet-isospin-triplet projectors $\mathcal{P}$ and $\mathcal{T}$ that select the corresponding $NN$ states are defined as
\begin{align}
    \mathcal{P}_i = \frac{1}{\sqrt{8}}\sigma_i\,,\qquad \mathcal{T}_a = \frac{1}{\sqrt{8}}\tau_a\,,
\end{align}
with the normalisation
\begin{align}
    \tr \mathcal{P}_i\mathcal{P}_j^\dagger = \frac{1}{2}\delta_{ij}\,, \quad  \tr \mathcal{T}_a\mathcal{T}_b^\dagger = \frac{1}{2}\delta_{ab}\,, \quad  \tr \mathcal{P}_i\mathcal{T}_b^\dagger = 0\,,
\end{align}
where the trace is over both the spin and isospin indices.
The quadratic and quartic Galilean-invariant combinations of the nucleon gauge derivatives and projectors are defined as
\begin{align}
\mathcal{O}_i^{(2)} &= \frac{1}{4}\left[
\overleftarrow{\bv{D}}^2\mathcal{P}_i -2 \overleftarrow{{ D}}_j\mathcal{P}_i\overrightarrow{{D}}_j + \mathcal{P}_i\overrightarrow{\bv{D}}^2
\right]
\,,
\label{eq:O2}
\\
\mathcal{O}_i^{(4)} &= \frac{1}{16}\left[
\overleftarrow{\bv{D}}^4 \mathcal{P}_i - 4\overleftarrow{\bv{D}}^2\overleftarrow{{D}}_j\mathcal{P}_i\overrightarrow{{D}}_j
+4\overleftarrow{{D}}_j\overleftarrow{{D}}_k \mathcal{P}_i \overrightarrow{{D}}_k \overrightarrow{{D}}_j +2\overleftarrow{\bv{D}}^2\mathcal{P}_i\overrightarrow{\bv{D}}^2
-4 \overleftarrow{{D}}_j\mathcal{P}_i \overrightarrow{{D}}_j\overrightarrow{\bv{D}}^2
+\mathcal{P}_i \overrightarrow{\bv{D}}^4
\right]\,
\label{eq:O4}
\end{align}
for the triplet $NN$ channel, and
\begin{align}
\mathcal{O}_i^{(2,s)} &= \frac{1}{4}\left[
\overleftarrow{\bv{D}}^2\mathcal{T}_i -2 \overleftarrow{{D}}_j\mathcal{T}_i\overrightarrow{{D}}_j + \mathcal{T}_i\overrightarrow{\bv{D}}^2
\right]
\,
\label{eq:02s}
\end{align}
for the singlet $NN$ channel. The arrows point in the direction of operation of the corresponding derivative.
Note that there is a different definition of $\mathcal{O}^{(4)}_i$ in literature, e.g., Refs.~\cite{Chen:1999bg,Rupak:1999rk} use
\begin{align}
\mathcal{O}_i^{(4)} &= \frac{1}{16}\left[
\overleftarrow{\bv{D}}^4 \mathcal{P}_i - 4\overleftarrow{\bv{D}}^2\overleftarrow{{D}}_j\mathcal{P}_i\overrightarrow{{D}}_j
+6\overleftarrow{\bv{D}}^2\mathcal{P}_i\overrightarrow{\bv{D}}^2
-4 \overleftarrow{{D}}_j\mathcal{P}_i \overrightarrow{{D}}_j\overrightarrow{\bv{D}}^2
+\mathcal{P}_i \overrightarrow{\bv{D}}^4
\right]\,.
\label{eq:O4nG}
\end{align}
The latter expression, however, is potentially problematic, because it does not produce the desirable Galilean-invariant form of the matrix element:
\begin{align}
\left\langle \bv{p}_1'\, \bv{p}_2'\right|N^\dagger \mathcal{P}_i N_c\ N_c^\dagger \mathcal{O}_i^{(4)} N\left|\bv{p}_1\, \bv{p}_2\right\rangle \propto (\bv{p}_1-\bv{p}_2)^4\,.
\end{align}
Indeed, it is straightforward to see that
\begin{align}
    (\bv{p}_1-\bv{p}_2)^4 &= (\bv{p}_1^2-2\bv{p}_1\cdot\bv{p}_2+\bv{p}_2^2)^2 \nonumber\\
    & = \bv{p}_1^2-4\bv{p}_1^2\, \bv{p}_1\cdot\bv{p}_2 +4(\bv{p}_1\cdot \bv{p}_2)^2 + 2\bv{p}_1^2\bv{p}_2^2 - 4 \bv{p}_1\cdot\bv{p}_2\, \bv{p}_2^2+\bv{p}_2^4\,,
\end{align}
showing that the operator in Eq.~\eqref{eq:O4} produces the correct contractions of $\bv{p}_1$ and $\bv{p}_2$, whereas Eq.~\eqref{eq:O4nG} produces an expression that cannot be reduced to $(\bv{p}_1-\bv{p}_2)^4$ in the general case. This issue appears to also affect the sixth-power operator $\mathcal{O}^{(6)}_i$ defined in Ref.~\cite{Rupak:1999rk} (this operator starts contributing at N4LO and is therefore not relevant to our calculation). One has to note that Eq.~\eqref{eq:O4nG} gives the correct result in the center-of-mass frame, and, since it is a relatively high-order operator, the terms that spoil the Galilean invariance are probably rather small in a typical calculation; however, it is generally safer to use the expression given in Eq.~\eqref{eq:O4}.

For the $SD$ mixing $NN$ interaction, we consider the assignment where the corresponding coupling is $O(P^{-1})$~\cite{Rupak:1999rk,Chen:1999tn}, and, consequently, this interaction first appears at $O(P)$ in the $NN$ potential, which corresponds to NNLO.
However, the $SD$ mixing term being proportional to a $D$-wave $NN$ operator and there being no other $D$-waves up to N3LO, one needs two insertions of this term in order to receive a contribution to unpolarised VVCS. This demotes the leading $SD$ mixing contribution to $O(P^3)$ in the $NN$ potential, or N4LO, making it irrelevant to the present calculation.

In addition to the $S$-wave interactions, we also include the spin-triplet-isospin-triplet interactions in a $P$-wave, entering at N3LO:
\begin{align}
 \mathcal{L}^{NN}_P&=-\frac{1}{4}\left[
   C_{{}^{3\!}P_0}\delta_{ia}\delta_{jb}
 + C_{{}^{3\!}P_1}\left(\delta_{ij}\delta_{ab}-\delta_{ib}\delta_{ja}\right)
 +2C_{{}^{3\!}P_2}\left(\delta_{ij}\delta_{ab}+\delta_{ib}\delta_{ja}-\frac{2}{3}\delta_{ia}\delta_{jb}\right)
 \right] \nonumber\\
 &\times N^\dagger \mathcal{O}^{(1,P)}_{ia} N_c\ N_c^\dagger \mathcal{O}^{(1,P)}_{jb} N\,,
\end{align}
with 
\begin{align}
    \mathcal{O}^{(1,P)}_{ia} = \overleftarrow{D}_i P_a \tau_3 - \tau_3 P_a \overrightarrow{D}_i\,.
\end{align}

To complete the Lagrangian in the two-nucleon sector, one has to include two-nucleon contact interactions with the electromagnetic fields, whose couplings are not fixed by the $NN$ interaction. The contact terms needed for our calculation are
\begin{align}
\mathcal{L}^{NN\gamma} & = e L_1^{M1_V} \left[N^\dagger \mathcal{P}_i N_c\ N_c^\dagger\mathcal{T}_3 N + \text{H.c.}\right]\, B_i - 2i e L_2^{M1_S}\, \epsilon_{ijk} N^\dagger \mathcal{P}_i N_c\ N_c^\dagger \mathcal{P}_j N\, B_k \nonumber\\
& - \frac{e}{2} L_1^{E1_V}\left[N^\dagger\mathcal{O}_{ij}^{(1,P)} N_c\ N_c^\dagger \mathcal{P}_j N + \text{H.c.}\right] E_i
+\frac{e}{2} L_3^{E1_V}\left[N^\dagger\mathcal{O}_{ij}^{(1,P)} N_c\ N_c^\dagger \mathcal{O}^{(2)}_j N + \text{H.c.}\right] E_i \nonumber\\
& + eL_1^{C0_S} N^\dagger\mathcal{P}_i N_c\ N_c^\dagger \mathcal{P}_i N\, \left(\bv{\nabla}\cdot\bv{E}\right)
- \frac{e}{2}L_3^{C0_S}\left[N^\dagger\mathcal{P}_i N_c\ N_c^\dagger \mathcal{O}^{(2)}_i N + \text{H.c.}\right]\, \left(\bv{\nabla}\cdot\bv{E}\right)\,.
\end{align}
The first two magnetic interactions contribute to $f_T$ at $O(P)$, or NLO, whereas the four electric contact terms contribute to $f_L$ also at $O(P)$, or respective N3LO. Note that the convention in literature is to add to the $L_2^{M1_S}$ piece its Hermitian conjugate term, which, however, just gives the factor of $2$ that we write out explicitly.

\begin{table}[htb]
    \centering
    \begin{tabular}{l|l|l}
        $\alpha=e^2/4\pi\simeq 1/137.036$ & $\hbar c\simeq 197.327$~MeV\,fm & $m_\pi\simeq 139.570$~MeV\\ 
        \hline 
        $M\simeq 938.919$~MeV &  $\mu_p\simeq 2.793$ & $\mu_n\simeq -1.913$\\
        \hline 
        $r_p=0.84087(39)$~fm & $r_n^2=-0.1161(22)\text{ fm}^2$ & $M_d\simeq 1875.613$~MeV\\
        \hline
        $E_d=-2.224575(9)$~MeV~\cite{Epelbaum:2019kcf}  & $Z=1.6893(30)$~\cite{Epelbaum:2019kcf} & $w_2=0.389\text{ fm}^3$~\cite{Chen:1999bg} \\
        \hline
        $\gamma_s=-8.3208(15)$~MeV~\cite{Hackenburg:2006qd} & $r_s=2.750(18)$~fm~\cite{Hackenburg:2006qd}& $C_{{}^{3\!}P_J}=-1.49\text{ fm}^4$~\cite{Chen:1999bg}
    \end{tabular}
    \caption{Values of parameters entering the calculation of the VVCS amplitude.
    Unless referred to otherwise, the values are taken from the Particle Data Group listing~\cite{Zyla:2020zbs} (with the value of $r_p$ corresponding to the $\mu$H Lamb shift; note that we use $M_p\simeq 938.272$~MeV, rather than $M$, for the Darwin-Foldy correction). See the text for additional information, and also for the values of the two-nucleon electromagnetic couplings.}
    \label{tab:couplings}
\end{table}
The coupling constants of the single-nucleon sector are well known, and we show their values in Table~\ref{tab:couplings}, which also contains the values of the $NN$ parameters used by us.
Most of the parameters taken from the Particle Data Group listing~\cite{Zyla:2020zbs} are shown rounded and without the corresponding uncertainties, as the latter are negligibly small compared to the projected N3LO precision of $\sim 1\%$. The two exceptions here are the values of $r_p$ and $r_n^2$; we take their uncertainties into account, e.g., in the extraction of $l_1^{C0_S}$ in Sec.~\ref{sec:Deuteron_FF_N3LO}, even though the corresponding effect is small. The empirical values of $E_d$ and $Z$ are taken from the recent review~\cite{Epelbaum:2019kcf}, with $Z$ calculated from the $S$-wave asymptotic normalisation factor. While the value of $Z$ appears the most important source of uncertainty due to the input parameters, its quoted uncertainty is below $0.2\%$ and can thus also be neglected in the final error estimate, whose dominant source is the omitted higher-order terms in the \piEFT/ expansion. The values of $w_2$ and $C_{3P_J}$, taken by us from Ref.~\cite{Chen:1999bg}, are given there without uncertainties, however, the effect of these constants is very small, so even an uncertainty of the order of 100\% would not noticeably change the numerical results we present.

The $S$-wave $NN$ coupling constants are fixed by reproducing the expansion of the $NN$ $T$-matrix in the triplet and singlet channels in \piEFT/. Throughout our calculation, we use the dimensional regularisation and the power divergence subtraction (PDS) scheme~\cite{Kaplan:1998tg,Kaplan:1998we} to regularise divergent loop integrals. This introduces a regularisation scale dependence into the calculation, for instance, the loop function that corresponds to a single loop in Fig.~\ref{fig:MLO} in this scheme is
\begin{align}
    I_0(E) & = i\int\frac{\mathrm{d}^4 l}{(2\pi)^4}\frac{1}{\left[l_0-\frac{\bv{l}^2}{2M}+i0\right]\left[E-l_0-\frac{\bv{l}^2}{2M}+i0\right]}\xrightarrow{\text{PDS}}
    -\frac{M}{4\pi}\left(\mu -\sqrt{-M E-i0}\right)\,,
\end{align}
where $E$ is the relative motion energy of the $NN$ pair, and $\mu$ the regularisation scale. The $NN$ coupling constants also depend on $\mu$ in such a way that the $T$-matrix is $\mu$-independent. Up to the order we are working at, the $NN$ coupling constants are expressed in terms of $\mu$, $\gamma$, $(Z-1)$, and $w_2$ in the triplet channel, and $\mu$, $\gamma_s$, and the singlet effective range $r_s$ in the singlet channel. Note that the constants in the triplet channel are expanded in powers of the expansion parameter as well:
\begin{align}
    C_{0} = C_0^{(-1)} + C_0^{(0)} + C_0^{(1)} + C_0^{(2)} + \dots\,,
\end{align}
and analogously for the other constants. This expansion is needed in order to keep the position of the deuteron pole and the value of the residue unchanged as higher-order corrections are included. We provide the expressions for the constants entering Eq.~\eqref{eq:LagNN} below for the sake of completeness. The triplet couplings read
\begin{align}
    C_{0}^{(-1)} & = - \frac{4\pi}{M}\frac{1}{\mu-\gamma}\,,\qquad
    C_0^{(0)} =\hpm \frac{2\pi}{M}\frac{(Z-1)\gamma}{(\mu-\gamma)^2}\,,\qquad
    C_0^{(1)} =\hpm \frac{\pi}{M}\frac{(Z-1)^2\gamma(\gamma-2\mu)}{(\mu-\gamma)^3}\,,\nonumber\\
    C_0^{(2)} &=\hpm \frac{\pi}{2M}\frac{(Z-1)^3\gamma(\gamma-2\mu)^2}{(\mu-\gamma)^4}
    +\frac{4\pi}{M}\frac{w_2\gamma^4}{(\mu-\gamma)^2}    \,, \nonumber \\
    C_{2}^{(-2)} & =\hpm \frac{2\pi}{M}\frac{(Z-1)}{\gamma(\mu-\gamma)^2}\,, \qquad
    C_2^{(-1)}=-\frac{2\pi}{M}\frac{(Z-1)^2\mu}{\gamma(\mu-\gamma)^3}\,, \nonumber \\
    C_2^{(0)}&=-\frac{\pi}{2M}\frac{(Z-1)^3(\gamma^2-4\mu^2)}{\gamma(\mu-\gamma)^4}
    +\frac{8\pi}{M}\frac{w_2\gamma^2}{(\mu-\gamma)^2}   \,,\nonumber \\
    C_4^{(-3)} &= -\frac{\pi}{M}\frac{(Z-1)^2}{\gamma^2(\mu-\gamma)^3}\,, \qquad
    S_4^{(-2)} =- \frac{\pi}{2M}\frac{(Z-1)^3(\gamma-4\mu)}{\gamma^2(\mu-\gamma)^4}+\frac{4\pi}{M}\frac{w_2}{(\mu-\gamma)^2}\,, \nonumber \\
    C_6^{(-4)} &=\hpm \frac{\pi}{2M}\frac{(Z-1)^3}{\gamma^3(\mu-\gamma)^4}\,.
\end{align}
Here, $S_4^{(-2)}=C_4^{(-2)} + \tilde{C}_4^{(-2)}$ is the only linear combination of $C_4^{(-2)}$ and $\tilde{C}_4^{(-2)}$ that contributes to $NN$ scattering. We note that for the following consideration, it is convenient to eliminate the coupling $C_4^{(-2)}$, expressing it via $S_4^{(-2)}$ and $\tilde{C}_4^{(-2)}$. The LO, NLO, and NNLO couplings given here coincide with those given in Ref.~\cite{Phillips:1999hh}, however, we are not aware of any of the N3LO couplings explicitly appearing in the literature.
 Finally, the singlet couplings are
\begin{align}
 C_0^s =-\frac{4\pi}{M}\frac{1}{\mu - \gamma_s} \,, \qquad C_2^s = \frac{2\pi}{M}\frac{r_s}{(\mu-\gamma_s)^2}\,,
\end{align}
and the $P$-waves couplings contribute in a single linear combination
\begin{align}
    C_{{}^{3\!}P_J}&= C_{{}^{3\!}P_0}+2 C_{{}^{3\!}P_1}+\frac{20}{3} C_{{}^{3\!}P_2}\,,
\end{align}
whose value, extracted in Ref.~\cite{Chen:1999bg} from the Nijmegen partial-wave analysis~\cite{Stoks:1993tb,Stoks:1994wp}, is shown in Table~\ref{tab:couplings}.

The couplings of the contact terms entering $\mathcal{L}^{NN\gamma}$ cannot be inferred from the parameters of the $NN$ scattering amplitude; these contact terms, however, are crucial in compensating the $\mu$ dependence of the VVCS amplitude, and information on their couplings can be obtained from the renormalisation group (RG) equations and from processes involving external electromagnetic probes. We discuss the role played by these contact terms as well as their determination in detail below in Sec.~\ref{sec:VVCS_results}.

\subsection{LSZ reduction}
\begin{figure}[htb]
    \centering
    \includegraphics[width=0.6\textwidth]{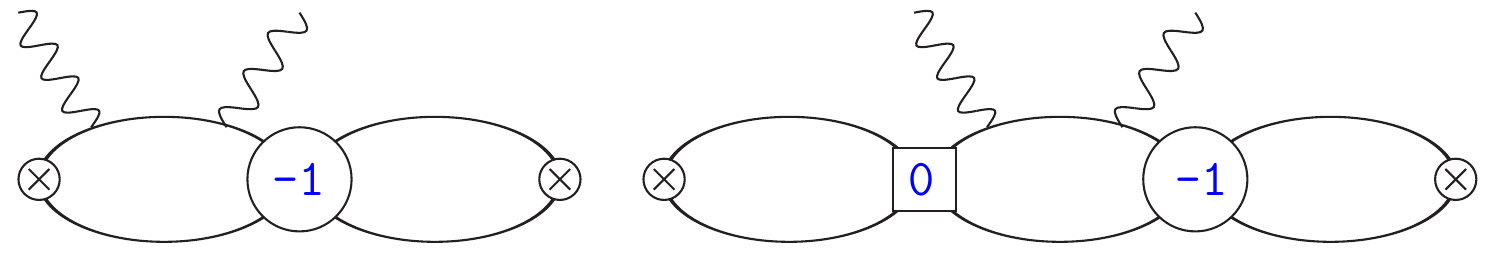}
    \caption{Examples of diagrams that have an external loop attached via an insertion of the LO triplet $NN$ $T$-matrix. Such diagrams are \emph{not} included in the four-point function. The crossed vertex denotes the leading-order $O(P^0)$ $NNd$ coupling.}
    \label{fig:reducible}
\end{figure}

In order to calculate the VVCS amplitude, we follow the LSZ reduction procedure along the lines described in, e.g., Ref.~\cite{Kaplan:1998sz} for the deuteron electromagnetic form factors. We introduce the coupling of the $NN$ system to an interpolating deuteron field that has the appropriate quantum numbers:
\begin{align}
    \delta\mathcal{L}^{NNd}= N^\dagger \mathcal{P}_i N_c\, \mathcal{E}_i+\text{H.c.}\,, 
\end{align}
where $\mathcal{E}_i$ are the spatial components of the deuteron polarisation vector (whose zeroth component vanishes in the deuteron rest frame). The VVCS amplitude is expressed via the sum of all four-point ($\gamma d\to \gamma d$) functions $\mathcal{M}(q,p,q',p')$ divided by the derivative of the deuteron self-energy $\Sigma(E)$ taken at the deuteron pole:
\begin{align}
    T_{fi} = \frac{\mathcal{M}(q,p,q',p')}{\Sigma^\prime(E_d)}\,.
    \label{eq:LSZ}
\end{align}
Diagrams that contribute to the four-point functions $\mathcal{M}$ are selected similarly to the calculation of the deuteron form factors, namely, only diagrams that do not have any external (incoming or outgoing) $NN$ loops attached via an insertion of the LO triplet $NN$ potential (or, equivalently, the LO triplet $NN$ $T$-matrix, see Fig.~\ref{fig:MLO}) contribute to $\mathcal{M}$. Diagrams that do have such external $NN$ loops attached, on the other hand, contribute to the dressing of the $NNd$ vertex, and are taken into account by the factor $\left[\Sigma'(E_d)\right]^{-1}$ in Eq.~\eqref{eq:LSZ}. 
Examples of such VVCS diagrams are shown in Fig.~\ref{fig:reducible}.
The self energy, in turn, is defined as the sum of all deuteron-deuteron two-point functions without any insertions of the LO $NN$ $T$-matrix. 
Note that by virtue of Eq.~\eqref{eq:LSZ} the normalisation of the interpolating deuteron field in $\delta\mathcal{L}^{NNd}$ is arbitrary. To obtain the order-by-order results, both $\mathcal{M}$ and $\Sigma^\prime(E_d)$ have to be expanded, and their ratio has to be expanded after that as well:
\begin{align}
    \mathcal{M}_L(\nu,Q^2) &= \mathcal{M}_L^{(-3)}+\mathcal{M}_L^{(-2)}+\dots\,,\\
    \mathcal{M}_T(\nu,Q^2) &= \mathcal{M}_T^{(-1)}+\mathcal{M}_T^{(0)}+\dots\,,\\
    \Sigma'(E_d) & = \Sigma^{\prime(-1)}+\Sigma^{\prime(0)}+\dots\,, \\
    f_L(\nu,Q^2) & = \underbrace{\vphantom{\Bigg[}\frac{\mathcal{M}_L^{(-3)}}{\Sigma^{\prime(-1)}}}_{O(P^{-2})}+\underbrace{\vphantom{\Bigg[}\frac{\mathcal{M}_L^{(-2)}}{\Sigma^{\prime(-1)}}-\frac{\mathcal{M}_L^{(-3)}\Sigma^{\prime(0)}}{\left[\Sigma^{\prime(-1)}\right]^2}}_{O(P^{-1})}+\dots\,,\\
    f_T(\nu,Q^2) & = \underbrace{\vphantom{\Bigg[}\frac{\mathcal{M}_T^{(-1)}}{\Sigma^{\prime(-1)}}}_{O(P^{0})}+\underbrace{\vphantom{\Bigg[}\frac{\mathcal{M}_T^{(0)}}{\Sigma^{\prime(-1)}}-\frac{\mathcal{M}_T^{(-1)}\Sigma^{\prime(0)}}{\left[\Sigma^{\prime(-1)}\right]^2}}_{O(P^{1})}+\dots\,,
\end{align}
where $\mathcal{M}_{L,T}(\nu,Q^2)$ denote the terms in $\mathcal{M}$ contributing to $f_{L,T}(\nu,Q^2)$.

\section{Diagrams and pertinent results, order by order}
\label{sec:VVCS_results}
In this section, we provide details of the order-by-order amplitudes necessary for the extraction of the observables.

\subsection{Deuteron self energy}
\begin{figure}[htb]
    \centering
    \includegraphics[width=\textwidth]{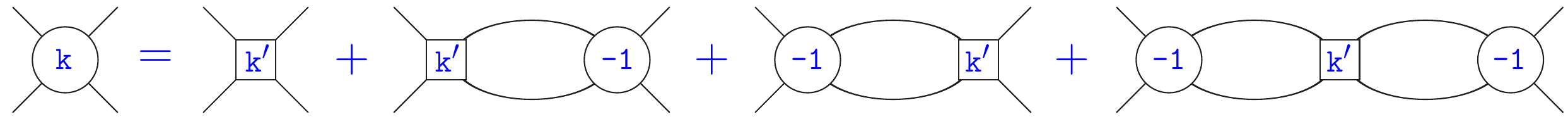}
    \caption{Correction to the $NN$ $T$-matrix vertex at order $k\ge 0$, defined in terms of the dressed potential at that order (denoted by the square with the primed index $k'$). Only the triplet channel is shown. The dressed potentials at the orders we are considering are shown in Fig.~\ref{fig:V_Dressed}.}
    \label{fig:Tmatrix}
\end{figure}

\begin{figure}[htb]
    \centering
    \includegraphics[width=\textwidth]{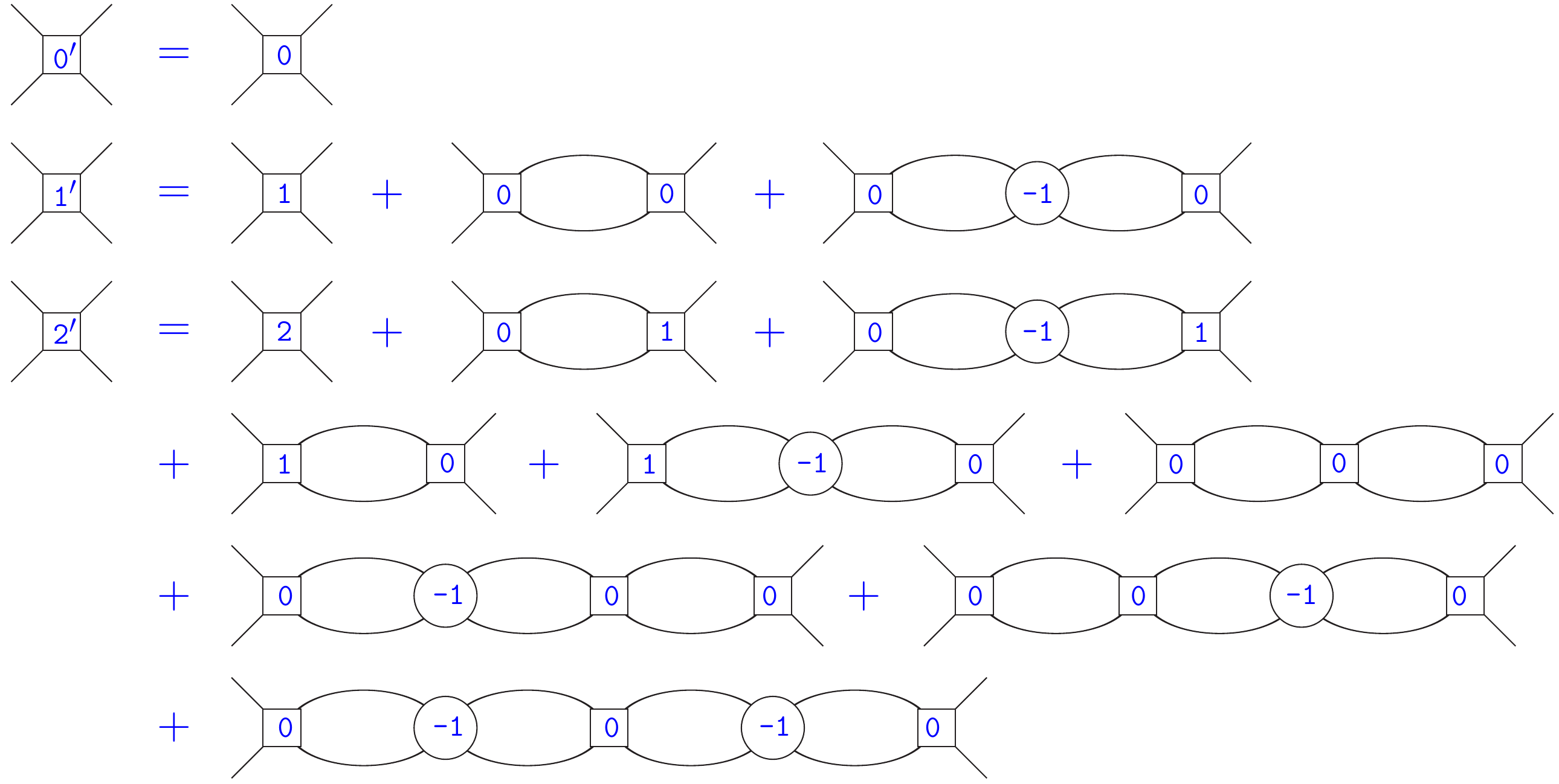}
    \caption{Dressed potential at $O(P^0)$, $O(P)$, and $O(P^2)$. Only the triplet channel is shown.}
    \label{fig:V_Dressed}
\end{figure}

\begin{figure}[htb]
    \centering
    \includegraphics[width=0.82\textwidth]{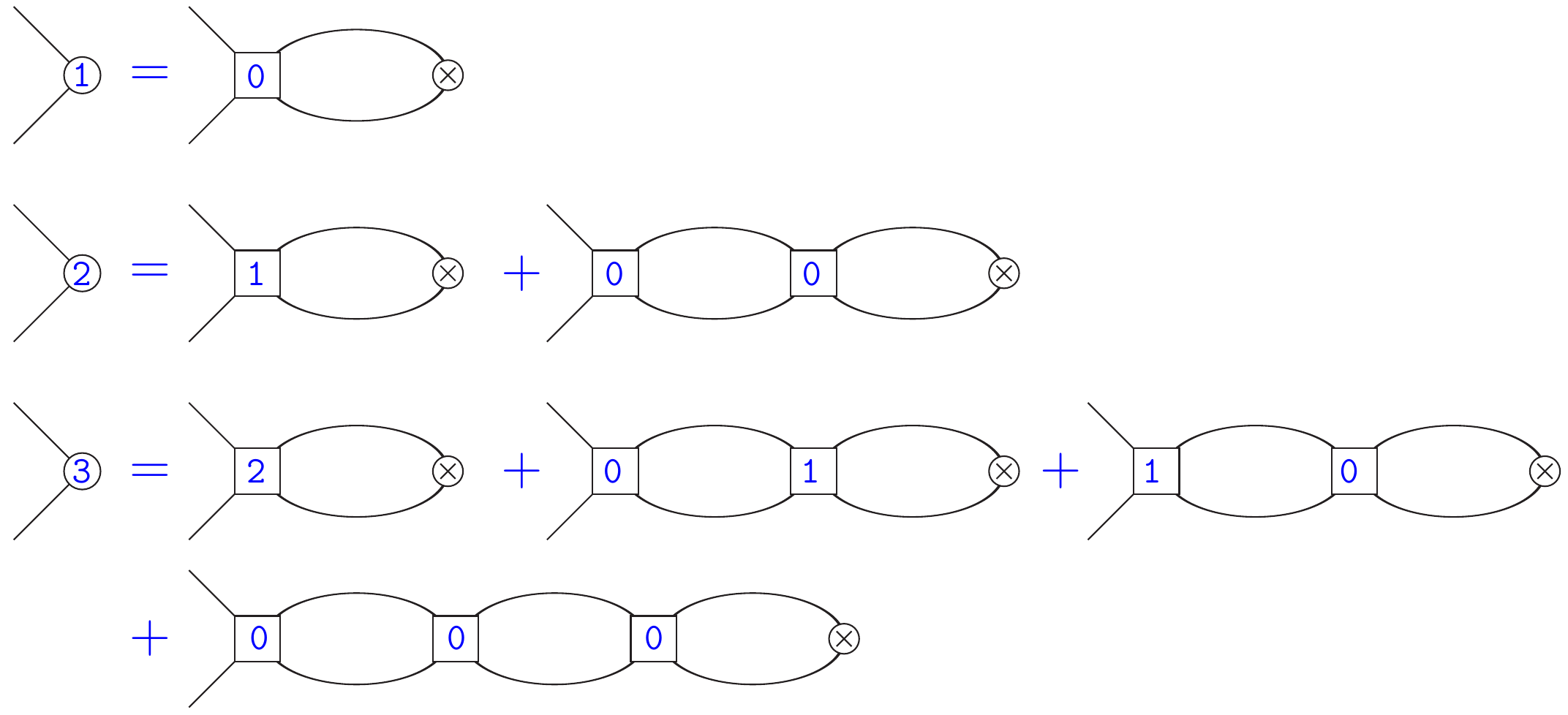}
    \caption{Corrections to the $NNd$ vertex at $O(P)$, $O(P^2)$, and $O(P^3)$, with the numbers denoting the order of the correction. Not to be confused with the $NN$ $T$ matrices defined in Figs.~\ref{fig:MLO} and~\ref{fig:Tmatrix}, which are denoted by bigger discs and have four external legs.}
    \label{fig:DeuteronVertex}
\end{figure}

To make the graphical representation of the matrix elements relevant for the calculation of the self energy $\Sigma(E)$ and the four-point function $\mathcal{M}(\nu,Q^2)$ more compact,
it is convenient to define certain subgraphs: the order-by-order corrections to the $NN$ $T$-matrix, shown in Fig.~\ref{fig:Tmatrix} and expressed via the dressed potentials that include all possible insertions of the LO $NN$ $T$-matrix in an intermediate state, shown in Fig.~\ref{fig:V_Dressed}; as well as the order-by-order corrections to the $NNd$ vertex, which are shown in Fig.~\ref{fig:DeuteronVertex}. Note that the $T$-matrix always appears in full off-shell kinematics.

\begin{figure}[htb]
    \centering
    \includegraphics[width=0.78\textwidth]{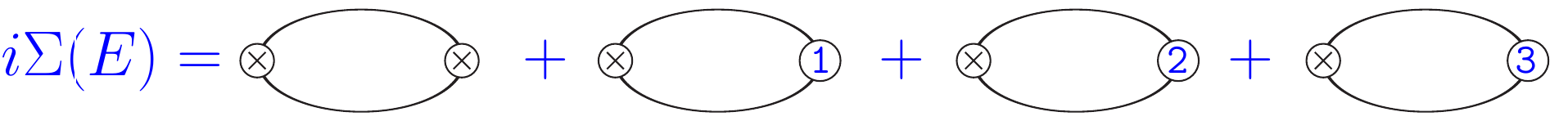}
    \caption{Graphical expression for $i\Sigma(E)$ up to N3LO.}
    \label{fig:Sigma}
\end{figure}

With these definitions, one obtains the very compact graphical form for $\Sigma(E)$ shown in Fig.~\ref{fig:Sigma}.
The resulting expression for $\Sigma(E)$ up to N3LO reads 
\begingroup
\allowdisplaybreaks[0]
\begin{align}
    \Sigma(E) = \frac{M}{4\pi}&\Bigg[\mu + ik \nonumber \\
    &+\frac{(Z-1) \left(\gamma ^2+k^2\right) (k-i \mu )^2}{2 \gamma  (\gamma -\mu )^2}\nonumber \\
    &+\frac{(Z-1)^2 \left(\gamma ^2+k^2\right) (\mu +i k)^2 \left(\gamma  \left(\gamma ^2-2 \gamma  \mu +2 \mu ^2\right)+i k^3-k^2 (\gamma -2 \mu )+i \gamma ^2 k\right)}{4 \gamma ^2 (\gamma -\mu )^4}\nonumber\\
    &-\frac{(Z-1)^3 \left(\gamma ^2+k^2\right) (k-i \mu )^2 \left(-i \gamma  \left(\gamma ^2-2 \gamma  \mu +2 \mu ^2\right)+k^3+i k^2 (\gamma -2 \mu
   )+\gamma ^2 k\right)^2}{8 \gamma ^3 (\gamma -\mu )^6}\nonumber\\
   &+\frac{w_2 \left(\gamma ^2+k^2\right)^2 (k-i \mu )^2}{(\gamma -\mu )^2}+\dots
    \Bigg]\,,
\end{align}
\endgroup
where $k=i\sqrt{-ME}$, and the N3LO result occupies the last two lines. The factors $\bv{\mathcal{E}}\cdot\bv{\mathcal{E}}^{\prime *}$ that correspond to the unpolarised deuteron are omitted. Even though $\Sigma(E)$ depends on $\mu$, its derivative at the deuteron pole, $\Sigma^\prime(E_d)$, is $\mu$-independent, and the order-by-order expression for it is very compact:
\begin{align}
\Sigma^\prime(E_d) &= \frac{M^2}{8\pi\gamma}\left[1-(Z-1)+(Z-1)^2-(Z-1)^3+\dots\right]\,,
\end{align}
giving an even simpler expression for the inverse quantity:
\begin{align}
    \left[\Sigma^\prime(E_d)\right]^{-1} & = \frac{8\pi\gamma}{M^2}\left[1+(Z-1)+0+0+\dots\right]\,.
\end{align}
As one could expect, the $Z$ factor in the residue is restored at NLO, and there are no corrections at higher orders.

\subsection{Four-point function}
Here, we present the order-by-order results for the four-point function $\mathcal{M}$.
\subsubsection{LO}
\begin{figure}[htb]
    \centering
    \includegraphics[width=\textwidth]{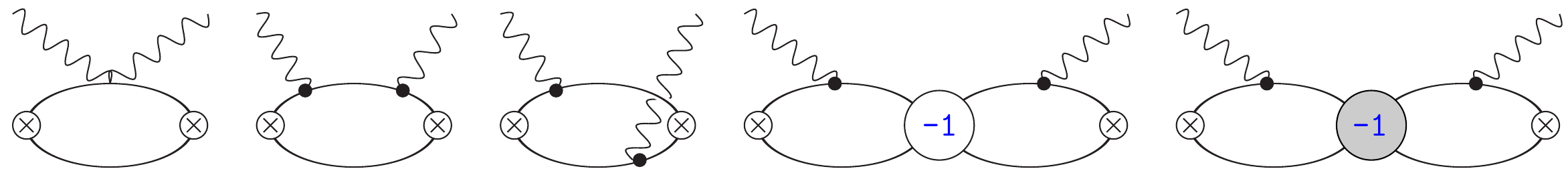}
    \caption{Diagrams contributing to $\mathcal{M}$ at LO, i.e., $O(P^{-3})$ for $\mathcal{M}_L$ and $O(P^{-1})$ for $\mathcal{M}_T$. Dotted vertices are the sum of the charge and the magnetic moment couplings from the single-nucleon Lagrangian; the big grey disc denotes the LO $NN$ $T$-matrix in the singlet channel. Crossed graphs are not shown. }
    \label{fig:LO}
\end{figure}
 The diagrams contributing to $i\mathcal{M}$ at LO are shown in Fig.~\ref{fig:LO}, resulting in
\begin{align}
\mathcal{M}_{L}^{(-3)} & = \frac{e^2M^3}{\pi}\frac{Q^2}{\bv{q}^2}\left[\frac{1}{2 \gamma  \left[\bv{q}^2+4 (\gamma +\lambda_d)^2\right]}-\frac{\phi^2(\nu,\bv{q}^2)}{\bv{q}^2 (\gamma-\lambda_d)}\right]+(\nu\to-\nu)\,,\\
\mathcal{M}_{T}^{(-1)} = \frac{e^2M}{\pi}\Bigg[& -\frac{16\gamma  (\gamma -\lambda_d) (\gamma +\lambda_d)^2+4 \bv{q}^2 \left(2 \gamma ^2+\gamma  \lambda_d+\lambda_d^2\right)-\left(4 \mu_1^2+4 \mu_0^2-1\right)\bv{q}^4 }{16 \gamma\,  \bv{q}^2
   \left[\bv{q}^2+4 (\gamma +\lambda_d)^2\right]}\nonumber\\
   &+
 \left(\frac{|\bv{q}| (\mu_0^2-\mu_1^2)}{12 M\nu}+\frac{4 \gamma ^2-4 \lambda_d^2+\bv{q}^2}{8 |\bv{q}|^3}\right)\phi(\nu,\bv{q}^2)\nonumber\\
  &
-\frac{1}{3} \left(\frac{2 \mu_0^2}{\gamma-\lambda_d}
+\frac{\mu_1^2}{\gamma_s-\lambda_d}\right)\phi^2(\nu,\bv{q}^2)
   \Bigg]+(\nu\to-\nu)\,,
\label{eq:MLO}
\end{align}
where the kinematic functions are defined as
\begin{equation}
\lambda_d=\sqrt{\gamma^2-M\nu+\frac{\bv{q}^2}{4}}\,,\qquad \phi(\nu,\bv{q}^2)=\arctan\frac{\left|\bv{q}\right|}{2(\gamma+\lambda_d)}\,.
\end{equation}
These expressions (as well as those at higher orders shown below) include the pole parts, with the elastic poles shifted due to the non-relativistic expansion and located at $\nu=\pm\bv{q}^2/(4M)$. One can also note that all loop integrals at LO are convergent, so no $\mu$ dependence emerges at this step. Another important remark regarding the set of graphs in Fig.~\ref{fig:LO}: together with the respective crossed graphs (not shown in the figure), the contribution of this set to the four-point function (and, correspondingly, to the VVCS amplitude) is gauge-invariant. This pattern is followed by the higher-order contributions presented below, where each figure shows a set of graphs that, together with the respective crossed counterparts, gives a gauge-invariant result. Note that some of the graphs shown may only serve to recover the gauge invariance and give a zero contribution to $\mathcal{M}$, however, we do not show graphs whose contribution vanishes identically (due to, e.g., isospin or dimensional regularisation factors).

\subsubsection{NLO}
\begin{figure}[htb]
    \centering
    \includegraphics[width=\textwidth]{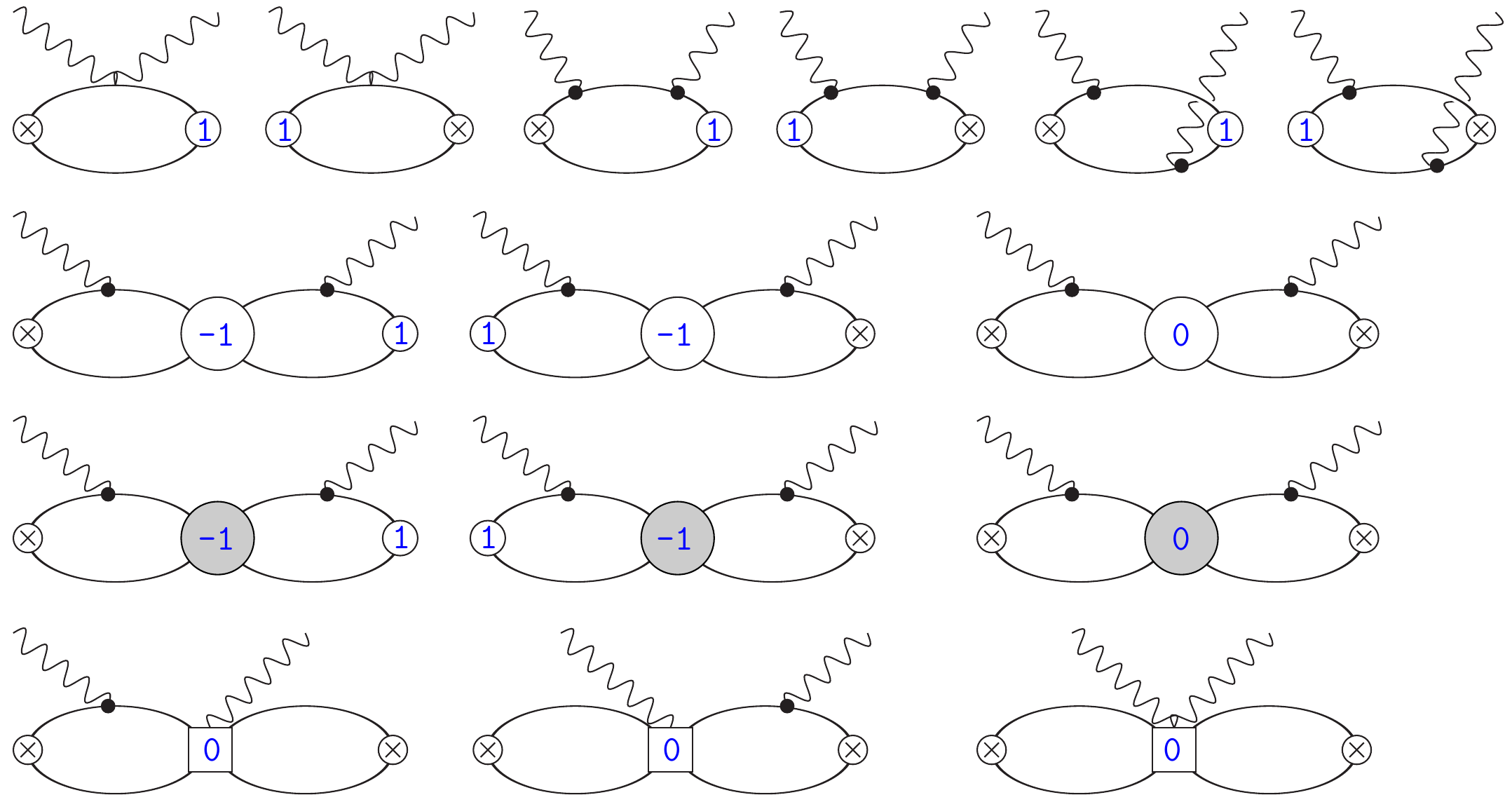}
    \caption{Contributions to $\mathcal{M}$ at NLO, i.e., $O(P^{-2})$ for $\mathcal{M}_L$ and $O(P^0)$ for $\mathcal{M}_T$, due to NLO corrections in the $NN$ interaction. The notation is as in Fig.~\ref{fig:LO}. Crossed graphs are not shown.}
    \label{fig:NLO}
\end{figure}

\begin{figure}[htb]
    \centering
    \includegraphics[width=0.8\textwidth]{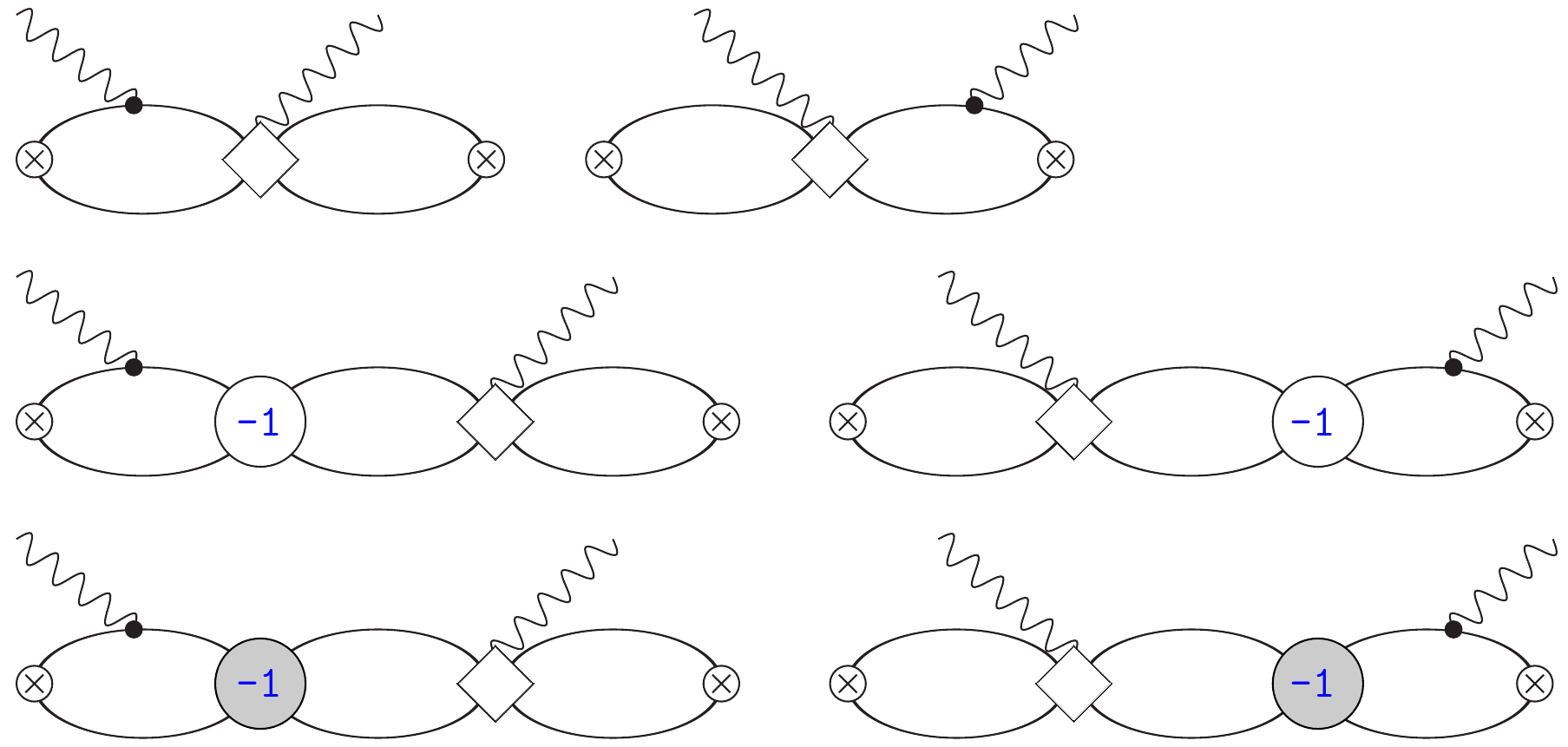}
    \caption{Contributions to $\mathcal{M}$ at NLO due to the magnetic contact terms $L_1^{M1_V}$ and $L_2^{M1_S}$ (shown as diamonds). The rest of the notation is as in Fig.~\ref{fig:LO}. These graphs give an $O(P^0)$ contribution to $\mathcal{M}_T$ only. Crossed graphs are not shown.}
    \label{fig:NLO_CT}
\end{figure}

The NLO contributions to $\mathcal{M}$ come from diagrams shown in Figs.~\ref{fig:NLO} and~\ref{fig:NLO_CT}.
While the contributions to $\mathcal{M}_{L}^{(-2)}$ only come from the graphs in Fig.~\ref{fig:NLO}, and their sum is $\mu$ independent, $\mathcal{M}_{T}^{(0)}$ also receives contributions from the
two magnetic contact terms in $\mathcal{L}^{NN\gamma}$, corresponding to the graphs in Fig.~\ref{fig:NLO_CT}. 
The renormalisation scale dependence of the coupling constants $L_1^{M1_V}$ and $L_2^{M1_S}$ has to cancel out the $\mu$ dependence of the total NLO contribution to $\mathcal{M}_{T}$. We find the following RG equations for these two couplings:
\begin{align}
\mu\frac{\mathrm{d}\hphantom{\mu}}{\mathrm{d}\mu}\left[(\mu-\gamma)(\mu-\gamma_s)\left(L_1^{M1_V}-\frac{\mu_1}{2}\left\{C_2^{(-2)}+C_2^{(-2,s)}\right\}\right)\right] & = 0\,,
\label{eq:RG_magnetic_1}
\\
\mu\frac{\mathrm{d}\hphantom{\mu}}{\mathrm{d}\mu}\left[\frac{L_2^{M1_S}}{C_2^{(-2)}}\right] & = 0\,.
\label{eq:RG_magnetic_2}
\end{align}
This is consistent with the results obtained in Refs.~\cite{Chen:1999bg,Rupak:1999rk,Ji:2003ia}. The first of these coupling constants, $L_1^{M1_V}$, contributes to the deuteron $\beta_{M1}$. The value of the RG-invariant quantity in Eq.~\eqref{eq:RG_magnetic_1} can be fitted to, e.g., data on $np\to d\gamma$; we use the result from Ref.~\cite{Rupak:1999rk} that found, also in the $z$-parametrisation,
\begin{equation}
(\mu-\gamma)(\mu-\gamma_s)\left(L_1^{M1_V}-\frac{\mu_1}{2}\left\{C_2^{(-2)}+C_2^{(-2,s)}\right\}\right) =-9.039(27)\text{ fm}^2\,.
\label{eq:L1M_value}
\end{equation}
The value of $L_2^{M1_S}$ is fitted to the deuteron magnetic moment, obtaining~\cite{Kaplan:1998sz,Chen:1999tn}
\begin{equation}
L_2^{M1_S}\Big|_{\mu=m_\pi}=-0.149\text{ fm}^4\,;
\label{eq:L2M_value}
\end{equation}
this result is the same  at NLO both in the $z$- and in the $\rho$-parametrisation.
To make the expression for $\mathcal{M}_T^{(0)}$ more compact, we introduce dimensionless $\mu$-independent couplings $l_1^{M1_V}$ and $l_2^{M1_S}$ according to
\begin{align}
L_1^{M1_V} & = \frac{\mu_1}{2}\left(C_2^{(-2)}+C_2^{(-2,s)}\right)+\frac{\pi (Z-1)}{M\gamma}\frac{l_1^{M1_V}}{(\mu-\gamma)(\mu-\gamma_s)}\,,\\
L_2^{M1_S} & = l_2^{M1_S}\, C_2^{(-2)}\,.
\end{align}
The resulting expressions for $\mathcal{M}_L$ and $\mathcal{M}_T$ read
\begin{align}
    \mathcal{M}_L^{(-2)} & =\frac{e^2 M^3}{\pi}\frac{Q^2}{\bv{q}^2} \frac{(Z-1)}{2\gamma}\frac{ \phi(\nu,\bv{q}^2) \left[|\bv{q}|-(\gamma +\lambda_d) \phi(\nu,\bv{q}^2)\right]}{ \bv{q}^2 (\gamma -\lambda_d)}+(\nu\to-\nu)\,,\\
    \mathcal{M}_T^{(0)}  = \frac{e^2 M}{\pi}\Bigg[&
    \frac{Z-1}{32 \gamma }
    -\frac{(Z-1)}{12\gamma}|\bv{q}|
    \left(
    \frac{\mu_1\,l_1^{M1_V} }{\gamma_s-\lambda_d}-\frac{4 \mu_0(\mu_0-2l_2^{M1_S})}{\gamma-\lambda_d}
    \right)\phi(\nu,\bv{q}^2)\nonumber\\
    &+\frac{1}{6} 
    \left(
    \frac{ \mu_1^2\, r_s\,\lambda_d^2}{(\gamma_s-\lambda_d)^2}-\frac{2 (Z-1) \mu_0^2 (\gamma +\lambda_d)}{\gamma  (\gamma -\lambda_d)}
    \right)\phi^2(\nu,\bv{q}^2)
   \Bigg]+(\nu\to-\nu)\,.
\end{align}
The values of the coupling constants $l_1^{M1_V}$ and $l_2^{M1_S}$, obtained using the values in Eqs.~\eqref{eq:L1M_value} and~\eqref{eq:L2M_value}, are
\begin{align}
l_1^{M1_V} = -4.596(14) \,,\qquad l_2^{M1_S} = -8.58\times 10^{-3}\,.
\label{eq:l1M_l2M_value}
\end{align}
While $l_1^{M1_V}$ could be considered as being of a natural size, $l_2^{M1_S}$ is numerically small due to the LO contribution already coming very close to the empirical value of the deuteron magnetic moment, as pointed out in Refs.~\cite{Kaplan:1998sz,Chen:1999tn}.
\subsubsection{NNLO}
\begin{figure}[htb]
    \centering
    \includegraphics[width=\textwidth]{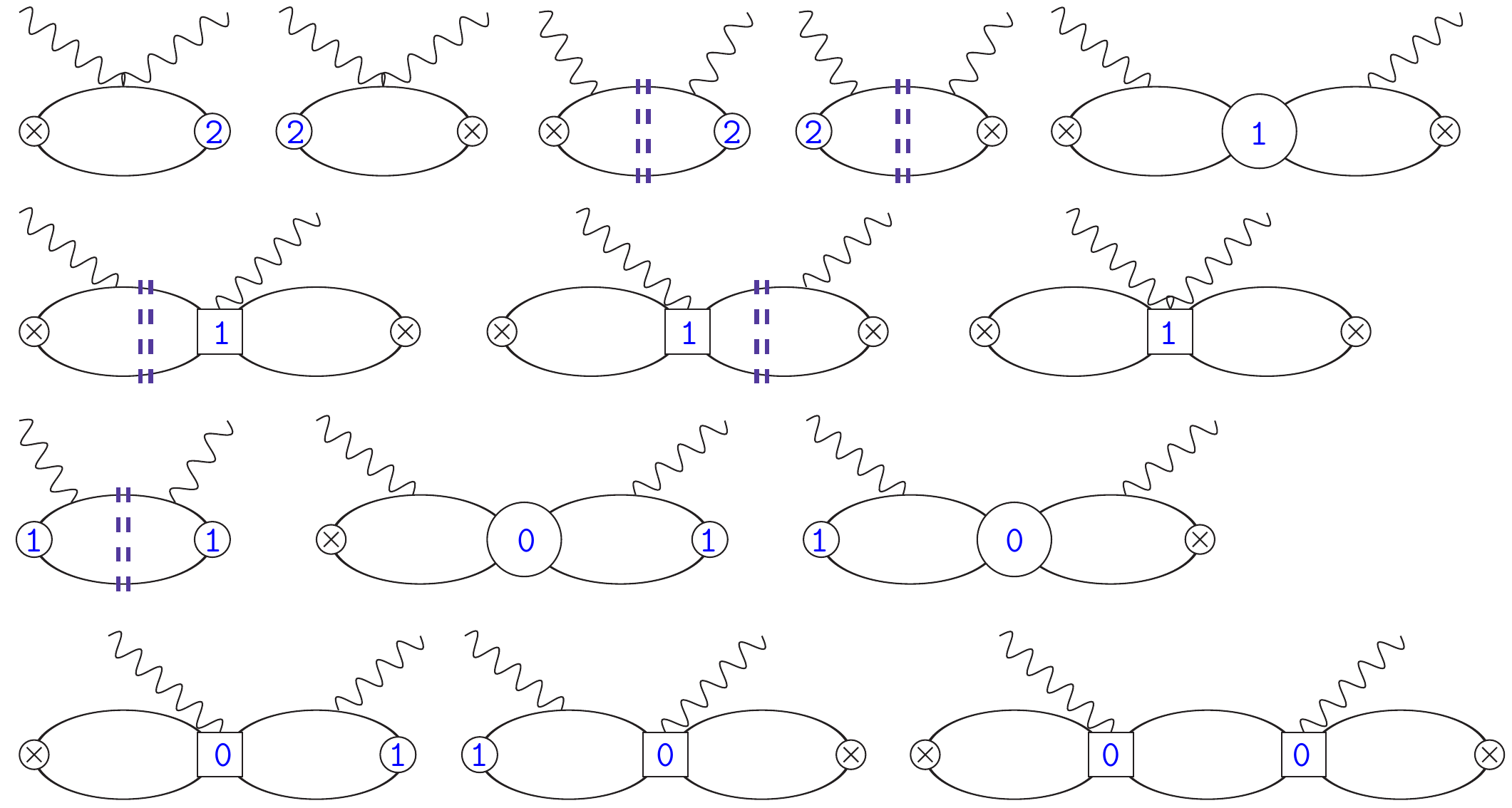}
    \caption{Diagrams that contribute to $\mathcal{M}_L$ at NNLO, i.e., $O(P^{-1})$, due to NNLO terms in the $NN$ interaction. The nucleon-photon vertex is exclusively the minimal coupling. In addition to diagrams that actually contribute to $f_L$, we also show those that are necessary in order to keep the electromagnetic gauge invariance. The vertical double dashed lines indicate possible insertions of a LO $NN$ $T$-matrix in the spin-triplet channel. Crossed graphs are not shown.}
    \label{fig:NNLO}
\end{figure}

\begin{figure}[htb]
    \centering
    \includegraphics[width=0.5\textwidth]{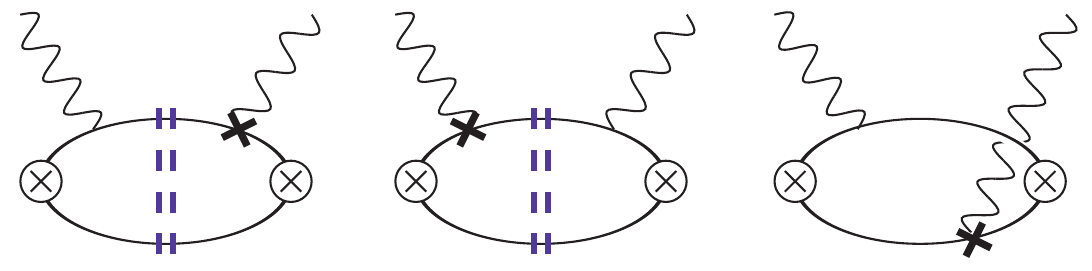}
    \caption{Correction to $\mathcal{M}_L$ due to the photon coupling proportional to $\hat{r}^2_E$, denoted by the black cross vertex, contributing at $O(P^{-1})$. The rest of the notation is as in Fig.~\ref{fig:NNLO}. Crossed graphs are not shown.}
    \label{fig:NNLO_RE}
\end{figure}

From this point on, we only keep track of those interactions that contribute to $f_L$, and, as a check, also include contributions to $\mathcal{M}$ that are needed to satisfy the electromagnetic gauge invariance. In practice this means taking into account the minimal coupling, both in the single-nucleon and in the two-nucleon Lagrangian. Incidentally, this allows one to also check that the Thomson term is recovered at NNLO and N3LO.
The corresponding diagrams that appear at NNLO are shown in Figs.~\ref{fig:NNLO} and~\ref{fig:NNLO_RE}.
At this order, one is still getting a $\mu$-independent contribution to $\mathcal{M}_L$ from loops with NNLO corrections to $NN$ interactions, Fig.~\ref{fig:NNLO}.
The resulting contribution to the longitudinal part of the four-point function is
 \begin{align}
\mathcal{M}_{L}^{(-1)} & =\frac{e^2M^3}{\pi}\frac{Q^2}{\bv{q}^2}\frac{(Z-1)^2}{4\gamma^2}
\left[-\frac{1}{4(\gamma -\lambda_d)}-\frac{ \phi(\nu,\bv{q}^2)}{|\bv{q}|}+
\frac{(\gamma +\lambda_d) \phi^2(\nu,\bv{q}^2)}{\bv{q}^2}
\right]+(\nu\to-\nu)\,.
 \end{align}
 In addition, corrections arise due to the nucleon charge radius operator, shown in Fig.~\ref{fig:NNLO_RE}. Their contribution is
\begin{align}
\delta\mathcal{M}_{L}^{(-1)}  =& -\frac{e^2 M^3}{6\pi}\frac{Q^2}{\bv{q}^2}\biggl[\frac{(r_0^2+r_1^2)\bv{q}^2 }{\gamma
\left[\bv{q}^2+4 (\gamma +\lambda_d )^2\right]}+
\frac{(r_0^2-r_1^2)|\bv{q}|  \phi(\nu,\bv{q}^2)}{M \nu }-\frac{4 r_0^2\, 
\phi^2(\nu,\bv{q}^2)}{\gamma -\lambda_d}\biggr]\nonumber\\
&+(\nu\to-\nu)\,.
\end{align}

\subsubsection{N3LO}
\begin{figure}[!htbp]
    \centering
    \includegraphics[width=0.98\textwidth]{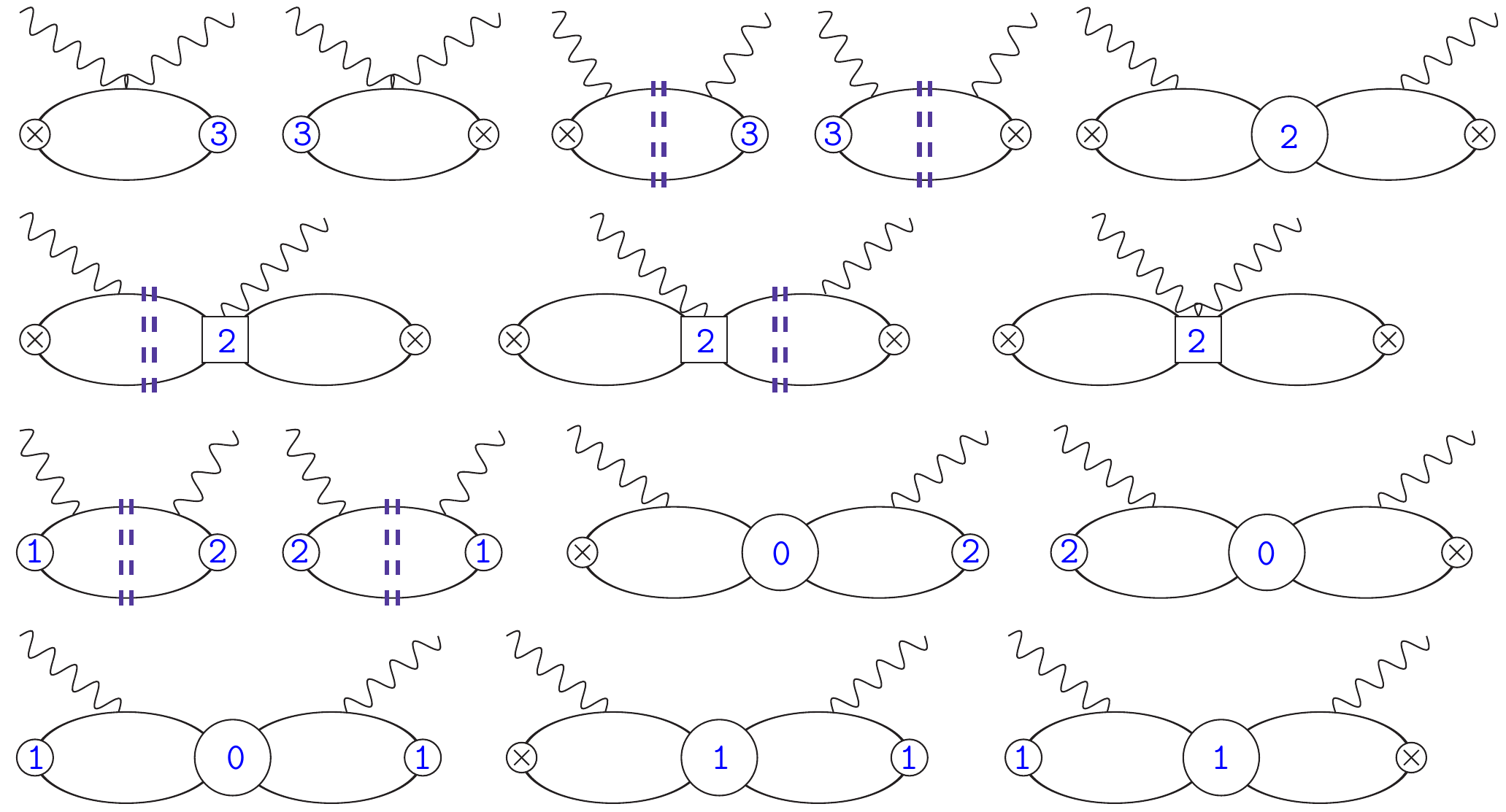}\\
    \includegraphics[width=\textwidth]{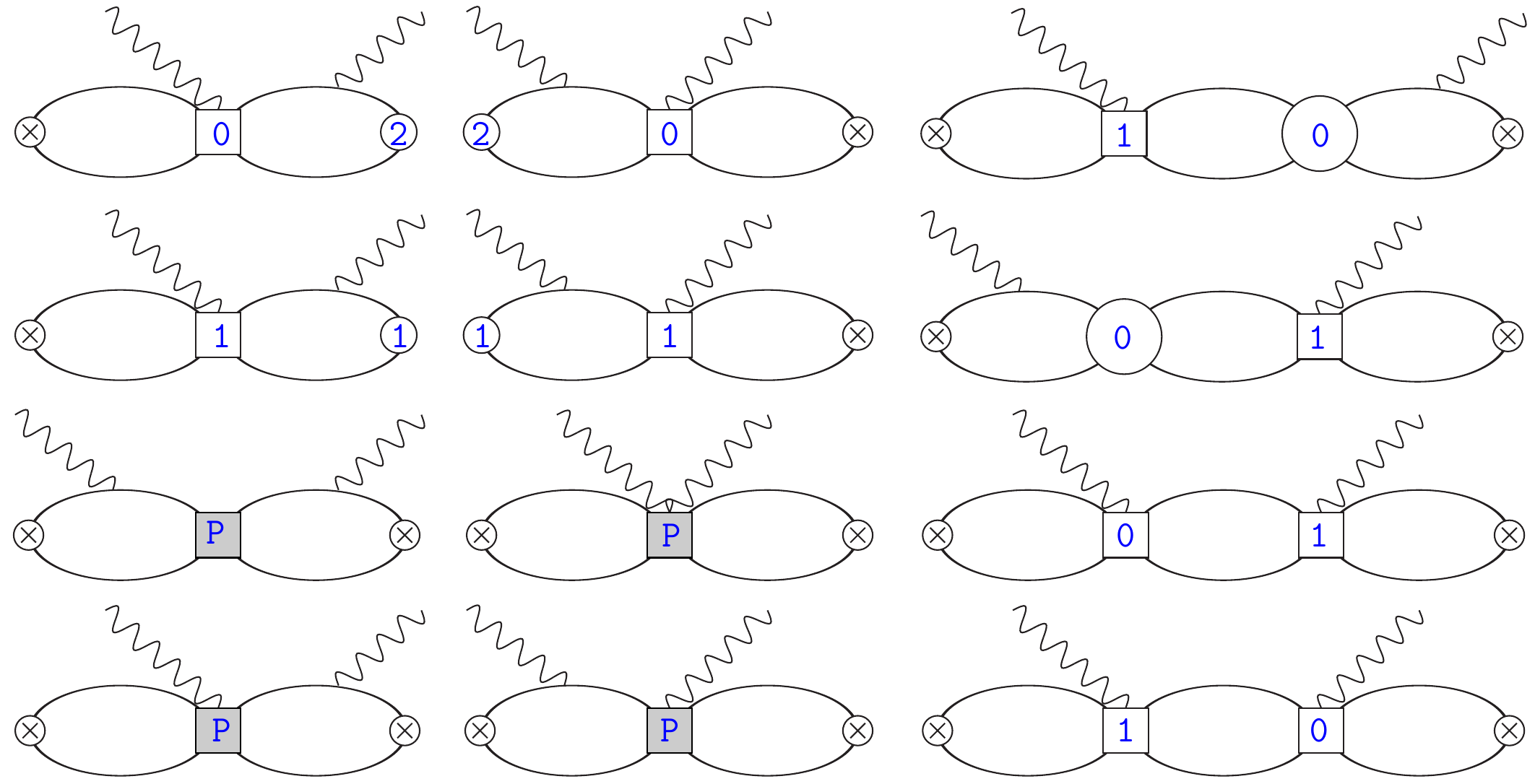}
    \caption{Diagrams that contribute to $\mathcal{M}_L$ at N3LO, i.e., $O(P^0)$, due to N3LO terms in the $NN$ interaction. In addition to diagrams that actually contribute to $f_L$, we also show those that are necessary in order to keep the electromagnetic gauge invariance. Grey squares marked by ``P'' show insertions of the $P$-wave $NN$ interactions. The rest of the notation is as in Fig.~\ref{fig:NNLO}. Crossed graphs are not shown.}
    \label{fig:NNNLO}
\end{figure}

\begin{figure}[htb]
    \centering
    \includegraphics[width=0.6\textwidth]{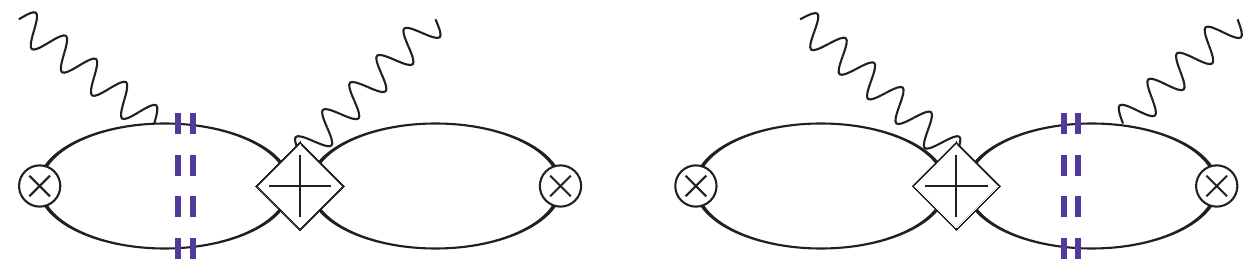}
    \caption{Contributions to $\mathcal{M}_L$ at N3LO due to the electric contact terms (shown as crossed diamonds). The rest of the notation is as in Fig.~\ref{fig:NNLO}. Crossed graphs are not shown.}
    \label{fig:NNNLO_CT}
\end{figure}

\begin{figure}[htb]
    \centering
    \includegraphics[width=0.7\textwidth]{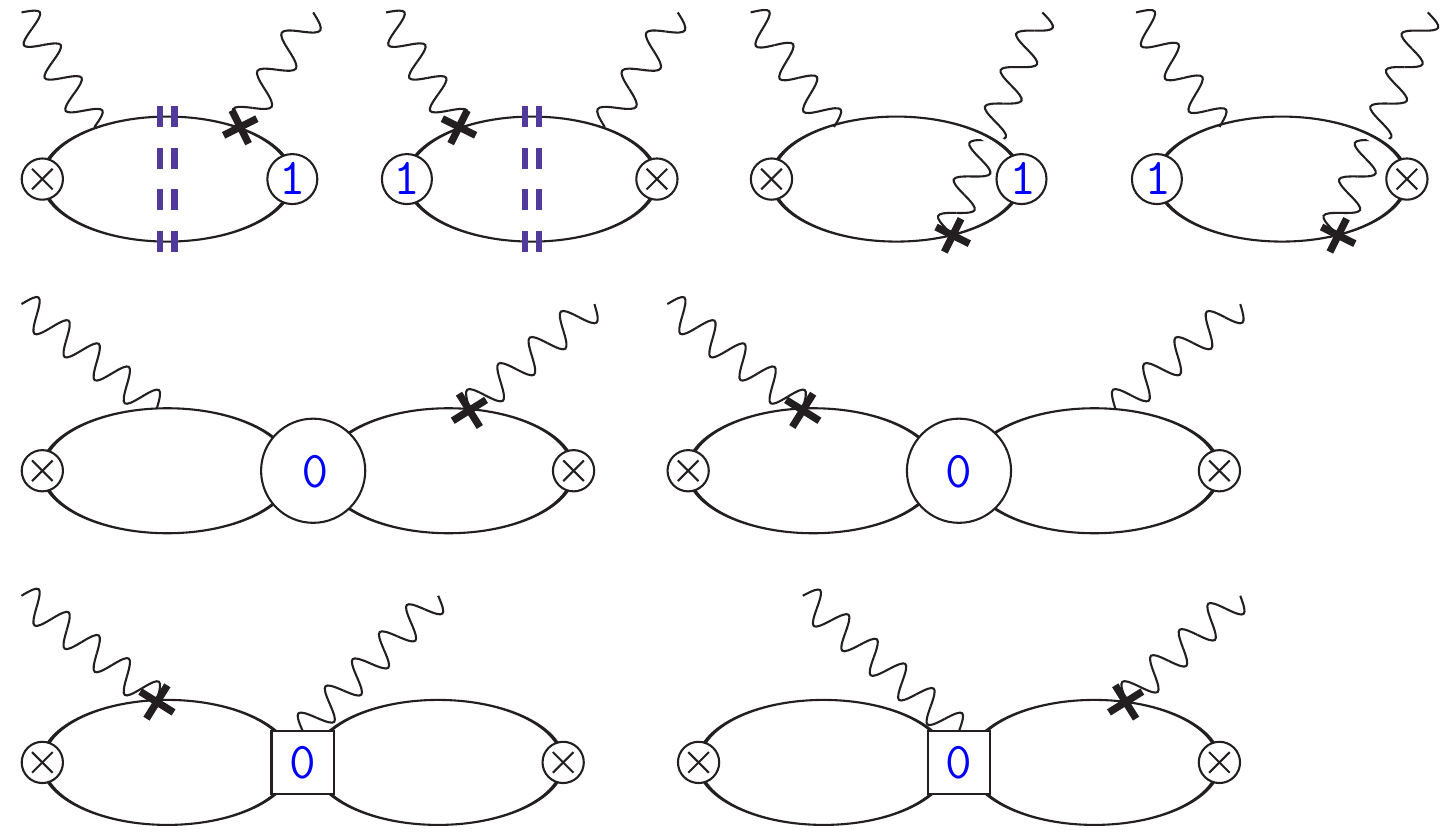}
    \caption{Correction to $\mathcal{M}_L$ due to the photon coupling proportional to $\hat{r}^2_E$, contributing at $O(P^0)$. The rest of the notation is as in Figs.~\ref{fig:NNLO} and~\ref{fig:NNLO_RE}. Crossed graphs are not shown.}
    \label{fig:NNNLO_RE}
\end{figure}

Finally, the N3LO contribution to $\mathcal{M}_L$ comes from three types of diagrams: those with insertions of the N3LO $NN$ interactions, shown in Fig.~\ref{fig:NNNLO}, diagrams with an insertion of a gauge-invariant electric contact term, shown in Fig.~\ref{fig:NNNLO_CT}, and corrections generated by one insertion of the nucleon charge radii coupling in the NLO graphs, as shown in Fig.~\ref{fig:NNNLO_RE}.
The diagrams in Fig.~\ref{fig:NNNLO} produce a $\mu$-dependent result, and all four electric contact terms are needed in order to render the total result RG invariant.
The couplings of these contact terms contribute to $\mathcal{M}_L^{(0)}$ in the following $\mu$-independent combinations with the $NN$ couplings:
\begin{align}
\mu\frac{\mathrm{d}\hphantom{\mu}}{\mathrm{d}\mu} \left[
\frac{L_1^{E1_V}-\frac{1}{2}M\, \tilde{C}_4^{(-2)}}{C_0^{(-1)}}
\right] & = 0\,,\qquad 
\mu\frac{\mathrm{d}\hphantom{\mu}}{\mathrm{d}\mu} \left[
\frac{L_3^{E1_V}-\frac{1}{2}M\, C_6^{(-4)}}{C_0^{(-1)}}
\right] = 0\,,\\
\mu\frac{\mathrm{d}\hphantom{\mu}}{\mathrm{d}\mu} \left[
\frac{L_1^{C0_S}+\frac{1}{4}M\,\tilde{C}_4^{(-2)}}{\left[C_0^{(-1)}\right]^2}
\right] & = 0\,,\qquad
\mu\frac{\mathrm{d}\hphantom{\mu}}{\mathrm{d}\mu} \left[
\frac{L_3^{C0_S}+\frac{1}{4}M\, C_6^{(-4)}}{\left[C_0^{(-1)}\right]^2}
\right] = 0\,.
\end{align}
The first two equations have been previously obtained in Ref.~\cite{Chen:1999bg};
our version, however, contains an additional factor $1/2$ in front of the nucleon mass in both equations, at variance with this reference.
The second pair of equations is new, to the best of our knowledge.

Considering the $\mu$ running of the quantities entering these RG equations at high momentum scales $\mu\simeq m_\pi$, one can conclude that
\begin{align}
L_1^{E1_V}-\frac{1}{2}M\, \tilde{C}_4^{(-2)}&=O(P^{-1}) \,,\qquad 
L_3^{E1_V}-\frac{1}{2}M\, C_6^{(-4)} = O(P^{-1})\,,\\
L_1^{C0_S}+\frac{1}{4}M\,\tilde{C}_4^{(-2)}&=O(P^{-2}) \,,\qquad
L_3^{C0_S}+\frac{1}{4}M\, C_6^{(-4)} = O(P^{-2})\,.
\end{align}
The first two combinations, being $O(P^{-1})$ instead of the na\"ively expected $O(P^{-2})$ and $O(P^{-4})$, are thus demoted to at least N4LO and N6LO, respectively. The same happens with the fourth combination, which is $O(P^{-2})$ instead of $O(P^{-4})$ and is demoted to N5LO. The only combination that gives a contribution at N3LO is the one that involves $L_1^{C0_S}$.
As we show in Sec.~\ref{sec:Deuteron_FF_N3LO}, its value can be found from a fit to the deuteron charge form factor.

Note that the cancellations between the contributions of the contact terms and those of the $NN$ couplings are in fact more intricate than given by these RG equations. The $NN$ coupling constants --- all apart from $\tilde{C}_4^{(-2)}$ --- conspire to remove the poles from the N3LO correction to the $NN$ $T$-matrix. The instances of $C_6^{(-4)}$ appearing in the RG equations are in fact combinations of all the $NN$ constants appearing at this order, so a statement that $L_3^{E1_V}$ and $L_3^{C0_S}$ cancel the contribution of $C_6^{(-4)}$ might be somewhat imprecise. The constant $\tilde{C}_4^{(-2)}$, on the other hand, is singled out from the other $NN$ constants, so its cancellations with the contact terms show a more transparent pattern. In particular, the cancellation is complete at N3LO in the transverse amplitude, consistent with what was shown previously in Refs.~\cite{Chen:1999bg,Rupak:1999rk} (even though our RG equations do not completely coincide with those references).

As before, we write
\begin{align}
    L_1^{C0_S} =-\frac{1}{4}M\, \tilde{C}_4^{(-2)} +  \frac{\pi(Z-1)^3}{\gamma^3(\mu-\gamma)^2}\,l_1^{C0_S}\,,
\end{align}
getting the following result for the total N3LO contribution of the diagrams in Figs.~\ref{fig:NNNLO} and~\ref{fig:NNNLO_CT}:
\begin{align}
    \mathcal{M}_L^{(0)} &=
    \frac{e^2M^3}{\pi}\frac{Q^2}{\bv{q}^2}\frac{(Z-1)^3}{\gamma^3}
    \Bigg[
    \frac{3 \gamma -\lambda_d}{2(\gamma -\lambda_d)}
    +\frac{2l_1^{C0_S}\, \bv{q}^2+(\gamma -\lambda_d)^2}{8 |\bv{q}|(\gamma-\lambda_d)}\phi(\nu,\bv{q}^2)
    -\frac{\gamma^2-\lambda_d^2 }{8 \gamma ^3 \bv{q}^2}\phi^2(\nu,\bv{q}^2)
    \Bigg]\nonumber \\
    &-\frac{e^2M^3}{\pi}\frac{Q^2}{\bv{q}^2}\frac{w_2 \left[|\bv{q}|-2 (\gamma +\lambda_d)\phi(\nu,\bv{q}^2)   \right]^2}{4 \bv{q}^2} \nonumber\\
    &+\frac{e^2M^3}{\pi}\frac{Q^2}{\bv{q}^2}
   \frac{M\,C_{{}^{3\!}P_J}\left[2 |\bv{q}| (\gamma-\lambda_d)-\left(4 \gamma ^2-4 \lambda_d^2+\bv{q}^2\right) \phi(\nu,\bv{q}^2)\right]^2}{192 \pi  \bv{q}^4}
    +(\nu\to-\nu)\,.
\end{align}
The nucleon charge radii corrections at this order, coming from the graphs in Fig.~\ref{fig:NNNLO_RE}, yield
\begin{align}
\delta\mathcal{M}_{L}^{(0)} & = -\frac{e^2 M^3}{3\pi}\frac{Q^2}{\bv{q}^2}
\frac{r_0^2 (Z-1)}{\gamma}\frac{\phi(\nu,\bv{q}^2)\left[|\bv{q}|-(\gamma +\lambda_d)\phi(\nu,\bv{q}^2)\right]}{\gamma-\lambda_d}+(\nu\to-\nu)\,.
\end{align}

Having established the order-by-order ingredients for the deuteron VVCS calculation throughout this section, we can now proceed with the extraction of the observables, form factors and generalised polarisabilities, in the following sections.

\section{Deuteron charge form factor at N3LO: fitting  $l_1^{C0_S}$}
\label{sec:Deuteron_FF_N3LO}
In order to extract the deuteron form factors, we use the residues of the pole parts of the VVCS amplitude. Using, e.g., the elastic structure functions~\cite{Carlson:2013xea}, one can obtain:
\begin{align}
    \res f_L(\nu,Q^2)\big|_{\nu=Q^2/(2M_d)} & = - e^2\left(1+\tau_d\right)
    \left[G_C^2(Q^2)+\frac{8}{9}\tau_d\,G_Q^2(Q^2)\right]\,,\\
    \res f_T(\nu,Q^2)\big|_{\nu=Q^2/(2M_d)} & = -\frac{2}{3}e^2 \tau_d\left(1+\tau_d\right) G_M^2(Q^2)\,,
\end{align}
where $G_C(Q^2)$, $G_M(Q^2)$, and $G_Q(Q^2)$ are the deuteron charge, magnetic, and quadrupole form factors, and $\tau_d=Q^2/(4M_d^2)$. At the order we are working, $1+\tau_d=1$, and $G_Q(Q^2)=0$. The shift of the elastic poles, which appear at $\nu=\pm\bv{q}^2/(4M)$, is also a relativistic correction that can be neglected at this order, allowing one to replace $\bv{q}^2\to Q^2$ in the amplitudes when taking the residues. This results in
\begin{align}
    \res f_L(\nu,Q^2)\big|_{\nu=Q^2/(4M)} & = - e^2\, G_C^2(Q^2)\,,\\
    \res f_T(\nu,Q^2)\big|_{\nu=Q^2/(4M)} & = -\frac{e^2}{24}\frac{Q^2}{M^2}\, G_M^2(Q^2)\,.
\end{align}
Evaluating the residues and expanding the square root of $G_C^2(Q^2)$ and $G_M^2(Q^2)$ order by order gives
\begin{align}
    G_C(Q^2) & =\frac{4 \gamma}{Q} \arctan\frac{Q}{4 \gamma } \nonumber\\
    & -(Z-1) \left(1-\frac{4 \gamma}{Q}\arctan\frac{Q}{4 \gamma}\right)\nonumber\\
    &-\frac{4}{3}r_0^2\, \gamma\,  Q  \arctan\frac{Q}{4 \gamma } \nonumber\\
    &+\frac{1}{3}(Z-1)\, r_0^2\,  Q^2\left(1-\frac{4 \gamma}{Q}\arctan\frac{Q}{4 \gamma }\right)-\frac{(Z-1)^3\,l_1^{C0_S}}{2 \gamma ^2} Q^2 \,, \label{eq:GC}\\
   \frac{e}{2M_d} G_M(Q^2) = \frac{e}{2M}\Bigg[& (\mu_n+\mu_p)\frac{4\gamma}{Q}\arctan\frac{Q}{4\gamma}\nonumber\\
    &-(\mu_n+\mu_p)(Z-1)\left(1-\frac{4\gamma}{Q}\arctan\frac{Q}{4\gamma}\right)
    +\frac{2}{\pi}M L_2^{M1_S}\gamma(\mu-\gamma)^2
    \Bigg]\,,\label{eq:GM}
\end{align}
at N3LO for $G_C(Q^2)$ and at NLO for $G_M(Q^2)$. The expression for $G_M(Q^2)$ coincides with that obtained in Ref.~\cite{Chen:1999tn} (where the $\gamma \rho_d$ factor in the NLO term has to be replaced by $(Z-1)$ to account for the difference between the $z$- and $\rho$-parametrisation schemes). The charge form factor has been previously calculated up to NNLO in the $z$-parametrisation in Ref.~\cite{Phillips:1999hh}, whose results we also reproduce here. Our expression for $G_C(Q^2)$ is also very similar to the result of Ref.~\cite{Ando:2004mm} that studied the deuteron form factor in the dibaryon formalism~\cite{Beane:2000fi}. Note, however, that Ref.~\cite{Ando:2004mm} appears to omit the Darwin-Foldy term.

The single-nucleon contributions can in principle be summed into the isoscalar nucleon form factors $G_{E,M}^{(p+n)}$, resulting in the following compact expressions for the deuteron form factors, valid up to N3LO for $G_C(Q^2)$ and up to NLO for $G_M(Q^2)$:
\bea
G_C(Q^2) & = & \frac{G_{E}^{(p+n)}(Q^2)}{\sqrt{1+\frac{Q^2}{4M_p^2}}}\left[\,Z\frac{4 \gamma}{Q} \arctan\frac{Q}{4 \gamma }-(Z-1)\right]  -\frac{(Z-1)^3\,l_1^{C0_S}}{2 \gamma ^2} Q^2 \,,  \label{eq:GC_summed}\\
G_M(Q^2) & = & \frac{M_d}{M} G_{M}^{(p+n)}(Q^2)\left[\, Z\frac{4 \gamma}{Q} \arctan\frac{Q}{4 \gamma }-(Z-1)\right]+4\frac{M_d}{M}(Z-1) l_2^{M1_S}\, \label{eq:GM_summed}.
\eea 
The square root in the denominator of Eq.~\eqref{eq:GC_summed} recovers the Darwin-Foldy term.
The order-by-order expression for the deuteron charge radius reads
\begin{align}
    R_C^2\equiv \big<r^2\big>_C = -6\frac{\mathrm{d}G_C(Q^2)}{\mathrm{d}Q^2}\bigg|_{Q^2=0} & =
    \frac{1}{8 \gamma ^2}
    +\frac{Z-1}{8 \gamma ^2}
    +2r_0^2
    +\frac{3(Z-1)^3}{\gamma ^2}\,l_1^{C0_S}\nonumber \\
    & = \left[2.3303+1.6063+0.6241+18.3166\, l_1^{C0_S} \right]\text{fm}^2\,.
    \label{eq:l1rc}
\end{align}
One has to note that the NNLO result $R_C^2=4.5607(76)\text{ fm}^2$ is already very close to the experimentally measured values, for instance, $\mu$D spectroscopy gives $R_C^2=4.5183(33)\text{ fm}^2$~\cite{Pohl1:2016xoo}. Using this value for fitting results in a tiny value of the $l_1^{C0_S}$ coupling,
\begin{align}
    l_1^{C0_S}=-2.32(18)(37)\times 10^{-3}\,,
\label{eq:contact_term_value}
\end{align}
similar to what happens with $l_2^{M1_S}$. The uncertainty in the first bracket is calculated using the quoted experimental error of the $\mu$D result. Considering other empirical values of $R_C^2$, e.g., the smaller value $R_C^2=4.3608\text{ fm}^2$ of the empirical parametrisation of Abbott et al.~\cite{Abbott:2000ak}, will significantly increase the uncertainty, albeit leaving $l_1^{C0_S}$ at a level of at most $10^{-2}$. The value in the second bracket is largely due to the uncertainty of the value of $Z$, shown in Table~\ref{tab:couplings}; one can see that the small relative uncertainty of $Z$ is amplified in the small subleading coupling $l_1^{C0_S}$. A small contribution to that value comes from the uncertainty of $r_0^2$; taken separately, the effects from $Z$ and from $r_0^2$ are, respectively, $0.35\times 10^{-3}$ and $0.13\times 10^{-3}$, and they are added in quadrature.

\begin{figure}[tb]
    \centering
    \includegraphics[width=0.55\textwidth]{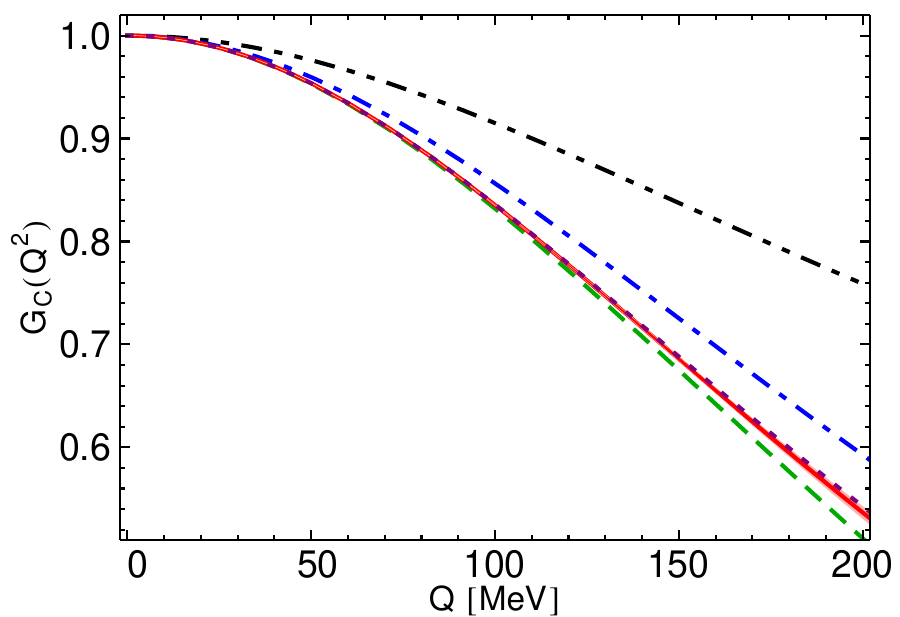}
    \caption{Deuteron charge form factor at LO (dash-dot-dotted black), NLO (dash-dotted blue), NNLO (dashed green), and N3LO (solid red, with the band showing the uncertainty due to higher-order terms).
    The result of the $\chi$EFT fit~\cite{Filin:2020tcs} is shown by the purple dotted curve.}
    \label{fig:GC}
\end{figure}

Figure~\ref{fig:GC} shows the deuteron charge form factor at the different orders, compared with
the result of the recent chiral EFT ($\chi$EFT) fit of Ref.~\cite{Filin:2020tcs} (see also Ref.~\cite{Filin:2019eoe}).
A conservative estimate for an error band at N3LO due to the higher-order terms could be obtained along the lines suggested in, e.g., Ref.~\cite{Epelbaum:2014sza}:
\begin{align}
\label{eq:errorGC}
\delta G_C(Q^2) = \max \left\{
\xi^4\Delta G_C^\mathrm{LO}(Q^2),\,
\xi^3\Delta G_C^\mathrm{NLO}(Q^2),\,
\xi^2\Delta G_C^\mathrm{NNLO}(Q^2),\,
\xi\Delta G_C^\mathrm{N3LO}(Q^2)
\right\}\,,
\end{align}
where $\xi=\gamma/m_\pi$ and $\Delta G_C(Q^2)$ are the contributions to $G_C(Q^2)$ at the respective orders, with terms $\propto Q^0,\ Q^2$ removed to account for the fact that $G_C(0)=1$, and $G_C'(0)=-\nicefrac{1}{6}R_C^2$ is also fixed at N3LO (the latter up to the small experimental error).
The resulting band is rather narrow, its width being about the difference between the \piEFT/ and $\chi$EFT results. An analogous band due to the uncertainty of $Z$ would in this scale be narrower than the widths of individual curves. We also do not show the theoretical uncertainty of the $\chi$EFT result; the corresponding band is roughly three times narrower than the \piEFT/ one and would not be visible in this scale.
One can see that the shapes of the charge form factor resulting from \piEFT/ and from $\chi$EFT agree very well even at values of $Q$ beyond the formal range of validity of \piEFT/ $Q\lesssim m_\pi$, up to photon virtualities $Q\lesssim 200$~MeV. This can be attributed to the fact that the slope and, to a lesser extent, the second derivative are by far the only important coefficients in the low-momentum expansion of $G_C(Q^2)$ in this range of $Q$, and they are well reproduced at N3LO in \piEFT/.

The observed agreement between \piEFT/ and $\chi$EFT at low $Q^2$ is very important. As low-energy effective field theories, both would be expected to provide a good description of the deuteron form factors at low $Q^2$ (at sufficiently high order in the respective expansion). This agreement vindicates the use of either theory as a tool to study the deuteron charge form factor at low $Q^2$. In practice, the simple analytic form of $G_C(Q^2)$ given in Eq.~\eqref{eq:GC} could be conveniently used to benchmark empirical parametrisations at low $Q^2$. One should nevertheless remember that the domain of validity of $\chi$EFT is considerably wider than that of \piEFT/, and that the $\chi$EFT calculation also has a considerably smaller theoretical uncertainty.

The recent empirical parametrisations of the charge form factor, such as those derived in Refs.~\cite{Abbott:2000ak,Sick:1998cvq} would lie in Fig.~\ref{fig:GC} roughly within a line thickness from the $\chi$EFT fit (or the N3LO \piEFT/ result), despite the numerical differences between the form factors (such as $G_C(Q^2)$ of Abbott et al.\ having a smaller value of $R_C^2$). We therefore do not show them here, either. Note, however, that these superficially small effects can have a rather sizeable influence on the values of the elastic contribution to the $\mu$D Lamb shift. This enhanced sensitivity can be used to judge on the quality of the empirical form factor parametrisations. One can also note that, since the value of the N3LO contact term $l_1^{C0_S}$ affects, through the value of $R_C^2$, the general shape of the deuteron charge form factor, one can deduce correlations between the value of $R_C^2$ and the elastic corrections. We shall consider these effects and compare the different variants of the deuteron charge form factor in detail in a dedicated publication~\cite{muDpaper}.

\section{Deuteron (generalised) polarisabilities}
\label{sec:Polarisabilities}
Having determined $l_1^{C0_S}$, we turn to the corresponding prediction for the deuteron polarisabilities. We start from the electric and magnetic dipole polarisabilities,
$\alpha_{E1}$ and $\beta_{M1}$, which can be read off from $f_L(\nu,Q^2)$ and $f_T(\nu,Q^2)$ using Eqs.~\eqref{eq:fLLEX} and~\eqref{eq:fTLEX}, respectively. The results are order-by-order given by
\begin{align}
    \alpha_{E1} & = \frac{\alpha M}{32\gamma^4}\left[1+(Z-1)+0+\frac{M\gamma^3}{6\pi}C_{{}^{3\!}P_J}\right] \nonumber \\
    & = [0.3771+0.2599+0-0.0018]\text{ fm}^3 = 0.6353\text{ fm}^3\,,\\
    \beta_{M1} & = \frac{\alpha}{32M\gamma^2}
    \bigg[-1+\frac{16}{3}\mu_1^2-\frac{32}{3}\mu_1^2\frac{\gamma}{\gamma_s-\gamma } \nonumber \\
&\hphantom{=\frac{\alpha}{32M\gamma^2}\bigg[\,\,}
+\frac{Z-1}{3} \left(16\mu_1^2-3\right)
-\frac{32(Z-1)}{3}\mu_1(\mu_1+l_1^{M1_V}) \frac{\gamma}{\gamma_s-\gamma}
+\frac{16}{3}\mu_1^2 r_s \frac{\gamma^3}{(\gamma_s-\gamma)^2}
   \bigg]\nonumber\\
& = [0.0701+0.0003]\text{ fm}^3 = 0.0704\text{ fm}^3\,,
\end{align}
at N3LO for $\alpha_{E1}$ and NLO for $\beta_{M1}$.
The expression for $\alpha_{E1}$ reproduces the N3LO result obtained in Ref.~\cite{Phillips:1999hh} using the $np\to\gamma d$ cross section calculated in Ref.~\cite{Rupak:1999rk}, however, our $P$-wave contribution is a factor of $2$ smaller. As a cross-check, we calculated the $P$-wave contribution to the $np\to d\gamma$ cross section, also getting a result twice smaller than obtained in Ref.~\cite{Rupak:1999rk}. The $P$-wave term, in any case, is very small numerically, making this disagreement insignificant in practice.
The numerical value is in agreement with, e.g., the recent evaluation of Ref.~\cite{Acharya:2020bxf} that obtained $\alpha_{E1}=0.626(18)\text{ fm}^3$ at N3LO in $\chi$EFT, as well as the calculation of Ref.~\cite{Hernandez:2014pwa} that used a selection of $\chi$EFT potentials with various cut-offs, along with the AV18 model potential~\cite{Wiringa:1994wb}. 

The LO expression for $\beta_{M1}$ reproduces the result of Ref.~\cite{Ji:2003ia}, and the NLO result is new to the best of our knowledge. The numerical value is also in a very good agreement with the N3LO $\chi$EFT result of Ref.~\cite{Acharya:2020bxf}, $\beta_{M1}=0.0715(15)\text{ fm}^3$; this agreement is remarkable, given that this is a relatively low-order calculation. One can notice that the contribution of the $l_1^{M1_V}$ contact term is individually rather sizeable $\simeq-0.06~\text{fm}^3$, and that it cancels almost completely with the remaining NLO terms.
This cancellation, giving essentially a zero NLO contribution to $\beta_{M1}$, appears even more surprising. While the agreement of $\beta_{M1}$ with the $\chi$EFT result is achieved already at LO and could be regarded accidental, a very small NLO contribution could be attributed to the procedure used in Ref.~\cite{Rupak:1999rk} to fit the value of $l_1^{M1_V}$ to reproduce the $np\to d\gamma$ cross section at NLO. Indeed, if this cross section is well described in \piEFT/, one should also expect a good description of the transverse response function of the deuteron at small non-zero $Q^2$. This, by virtue of the sum rule for $\beta_{M1}$ derived in Ref.~\cite{Gorchtein:2015eoa}, should be sufficient to reproduce the value of $\beta_{M1}$ at NLO, explaining at the same time the vanishing NLO contribution. This explanation, however, may imply that the description of the response function (or, more precisely, of its slope as function of $Q^2$ at $Q^2=0$) has to remain satisfactory up to relatively high energies outside the validity of \piEFT/; while this may be the case, investigating it in further detail is outside of the scope of this work. Nevertheless, the fact that the magnetic polarisability is reproduced accurately in \piEFT/ using input from $np\to d\gamma$ is a nice illustration of the working principles of an EFT, as well as a demonstration of the predictive power of the theory.

\begin{figure}[htb]
    \centering
    \begin{tabular}{cc}
    \includegraphics[width=0.491\textwidth]{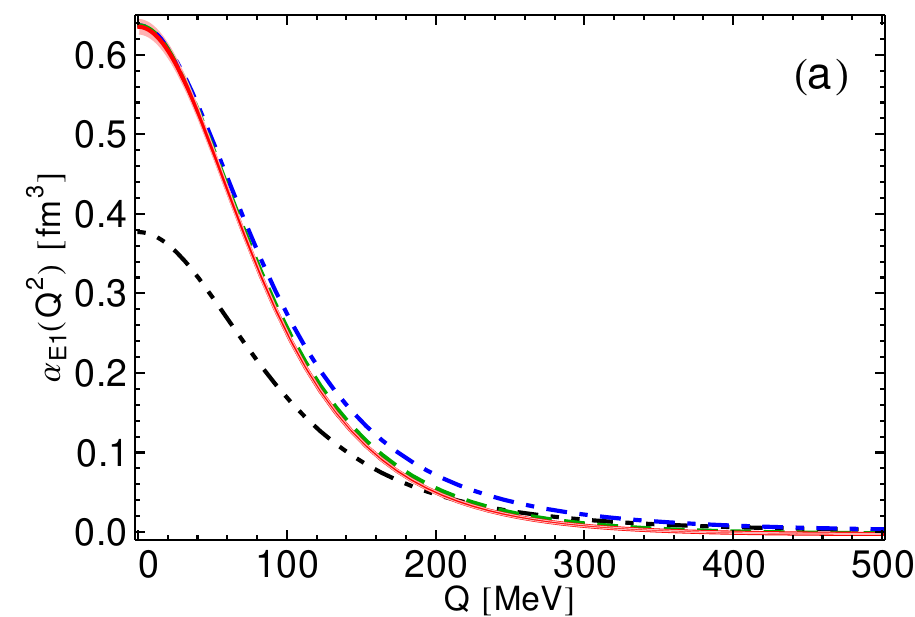} &
    \includegraphics[width=0.491\textwidth]{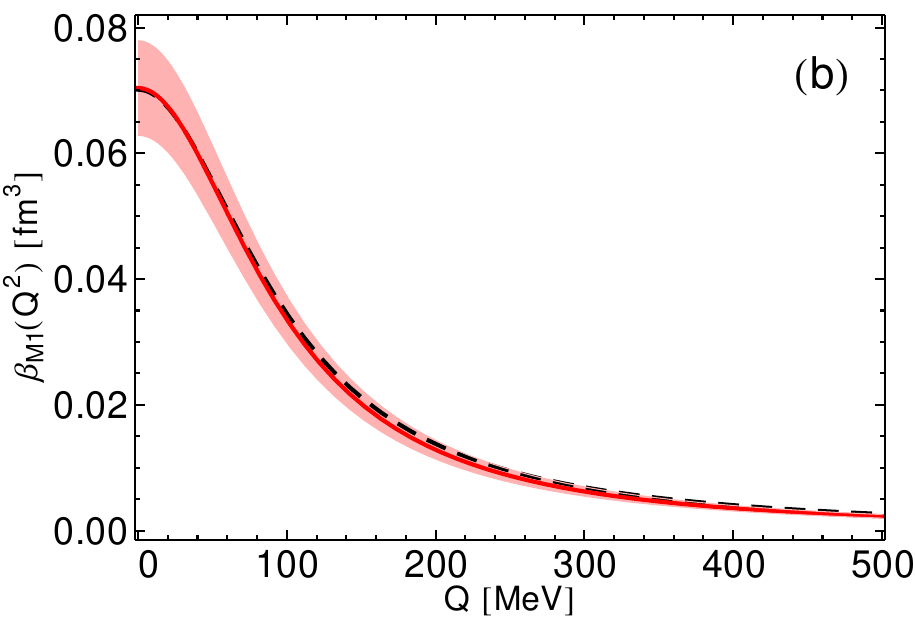}
    \end{tabular}
    \caption{Generalised deuteron polarisabilities: (a) $\alpha_{E1}(Q^2)$  and (b) $\beta_{M1}(Q^2)$. The LO, NLO, NNLO, and N3LO results for $\alpha_{E1}(Q^2)$ in the left panel are coded as in Fig.~\ref{fig:GC}. In the right panel, the LO and NLO results for $\beta_{M1}(Q^2)$ are shown, respectively, by the black dashed and the red solid curve, with the band showing the estimate of higher-order contributions.}
    \label{fig:alpha_beta}
\end{figure}

The generalisation of $\alpha_{E1}$ and $\beta_{M1}$ to finite $Q^2$ is defined in the usual way,
\begin{align}
    \alpha_{E1}(Q^2) = \frac{f_L(0,Q^2)}{4\pi Q^2}\,,\qquad 
    \beta_{M1}(Q^2)  = \frac{\bar{f}_T(0,Q^2)}{4\pi Q^2}\,,
\end{align}
where $f_L(0,Q^2)$ is understood as the non-pole part of $f_L$, and $\bar{f}_T$ stands for the non-pole part of $f_T$ with the Thomson term subtracted as well. The resulting curves are shown in Fig.~\ref{fig:alpha_beta}.
One can see that the patterns shown in the static values repeat in the generalised polarisabilities: the bulk of $\alpha_{E1}(Q^2)$ comes from the LO and NLO contributions (with a small but visible NNLO contribution that is mostly due to the nucleon charge radii corrections and vanishes at $Q=0$), whereas the NLO contribution to $\beta_{M1}(Q^2)$ is small at $Q=0$ as well as at finite virtualities.

The bands that estimate the contribution of higher orders are obtained analogously to Eq.~\eqref{eq:errorGC}, with the obvious modification for the NLO results.
The N3LO band on the $\alpha_{E1}$ curve is almost too narrow to be noticed, while the NLO band on $\beta_{M1}$ is expectedly much wider. This, in fact, might be an overestimation, especially at low values of $Q$, in view of the agreement between the NLO \piEFT/ result for $\beta_{M1}$ and the respective N3LO $\chi$EFT result. One can expect that the NLO value of $\beta_{M1}$ is therefore already close to its ``true'' value and will not change that much at higher orders. Indeed, according to Ref.~\cite{Rupak:1999rk}, one can fit the subleading contribution to $L_1^{M1_V}$ (appearing at NNLO) so as to keep describing the $np\to d\gamma$ cross section at NNLO (which would correspond to NNLO in $\beta_{M1}$). Extending the argument regarding the agreement of $\beta_{M1}$ with the $\chi$EFT result, one can expect the magnetic polarisability to keep its value at NNLO (up to possible small corrections, e.g., $\simeq 0.7\times 10^{-3}~\mathrm{fm}^3$ stemming from the single-nucleon magnetic polarisabilities). A more detailed investigation of this issue is, however, also outside of the scope of this work.

It is also interesting to consider two further generalised polarisabilities, namely, the longitudinal polarisability $\alpha_L(Q^2)$ and the generalised Baldin sum rule $[\alpha_{E1}+\beta_{M1}](Q^2)$, defined via the non-pole parts of the amplitudes as
\begin{align}
\alpha_L(Q^2) = \frac{1}{4\pi Q^2}\frac{\mathrm{d}f_L(\nu,Q^2)}{\mathrm{d}\nu^2}\bigg|_{\nu=0} \,, \qquad
[\alpha_{E1}+\beta_{M1}](Q^2)=\frac{1}{4\pi}\frac{\mathrm{d}f_T(\nu,Q^2)}{\mathrm{d}\nu^2}\bigg|_{\nu=0}\,.
\end{align}
The $Q^2=0$ value of $\alpha_L$ is given by
\begin{align}
    \alpha_L & = \frac{7\alpha M^3}{768\gamma^8}\left[1+(Z-1)+0+\frac{11M\gamma^3}{126\pi}C_{{}^{3\!}P_J}\right]\nonumber\\
    & = \left[0.865+0.597+0-0.002\right]\times 10^3\text{ fm}^5=1.460\times 10^3\text{ fm}^5\,.
\end{align}
For the sake of simplicity, this result is obtained by substituting $|\bv{q}|\to Q$ in the expressions for $f_L$ obtained above and thus neglects relativistic corrections of a relative size of roughly $0.2\%$. One can see that higher deuteron moments, such as $\alpha_L$, are numerically enhanced, unlike what happens in the case of the nucleon, see, e.g, Ref.~\cite{Alarcon:2020wjg}.

The $Q^2=0$ value of the Baldin sum rule coincides with the NLO result for $\alpha_{E1}+\beta_{M1}$ given above. Strictly speaking, this includes corrections beyond NLO, because $\beta_{M1}$ starts two orders higher than $\alpha_{E1}$, however, the explicit gauge invariance allows one to recover $\beta_{M1}$ by substituting $|\bv{q}|=\sqrt{Q^2+\nu^2}$ instead of neglecting $\nu$. Using $\bv{q}=Q$, on the other hand, changes the value of $[\alpha_{E1}+\beta_{M1}](Q^2)$ by about $10\%$ at $Q^2=0$ (by dropping the static value of $\beta_{M1}$). The difference quickly decreases with growing $Q$ and becomes negligible already at $Q\simeq 20$~MeV.
\begin{figure}[htb]
    \centering
    \begin{tabular}{cc}
    \includegraphics[width=0.491\textwidth]{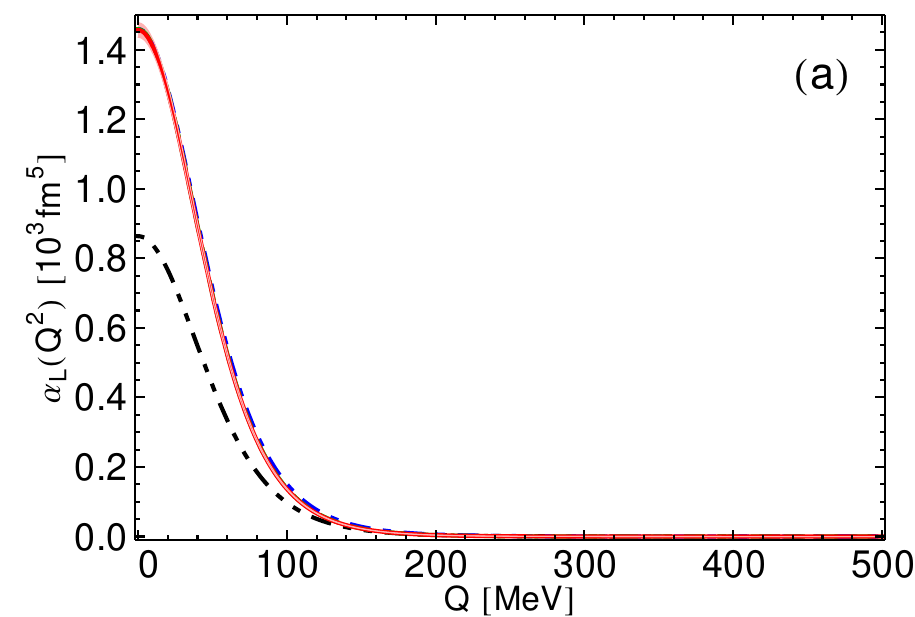} &
    \includegraphics[width=0.491\textwidth]{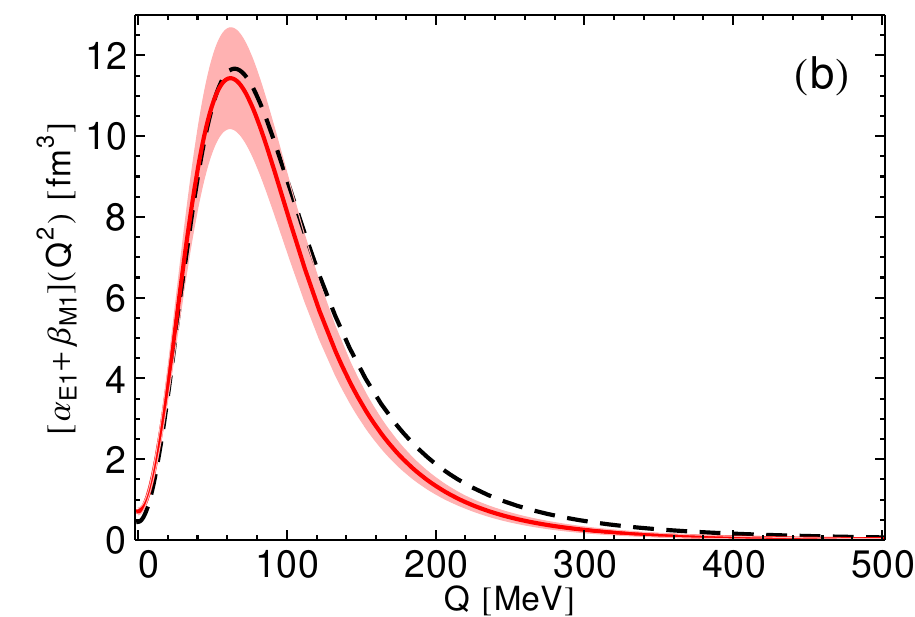}
    \end{tabular}
    \caption{Generalised deuteron polarisabilities: (a) $\alpha_{L}(Q^2)$ and (b) $[\alpha_{E1}+\beta_{M1}](Q^2)$. The curves are coded as in the left and right panel of Fig.~\ref{fig:alpha_beta}, respectively. The curves for $[\alpha_{E1}+\beta_{M1}](Q^2)$ are obtained using the exact expression $|\bv{q}|=\sqrt{\nu^2+Q^2}$ and reproduce at $Q=0$ the static values of $\alpha_{E1}+\beta_{M1}$ at the respective order, see the text for details.}
    \label{fig:alphaL_BSR}
\end{figure}

The curves for $\alpha_L(Q^2)$ and $[\alpha_{E1}+\beta_{M1}](Q^2)$ are shown in Fig.~\ref{fig:alphaL_BSR}. One can see that the longitudinal polarisability shares the general features of the previously considered $\alpha_{E1}(Q^2)$ and $\beta_{M1}(Q^2)$, with a somewhat quicker falloff with growing $Q$. The generalised Baldin sum rule, on the other hand, demonstrates a sharp increase peaking around $Q=60$~MeV; this enhancement is due to the magnetic interaction in the singlet channel and is analogous to what has been seen, e.g., in the generalised spin-forward deuteron polarisability $\gamma_0(Q^2)$~\cite{Lensky:2018vdq}.
The estimate of higher-order corrections to $\alpha_L(Q^2)$ and $[\alpha_{E1}+\beta_{M1}](Q^2)$, shown by the bands, is constructed analogously to, respectively, $\alpha_{E1}$ and $\beta_{M1}$. Similarly to those, the N3LO band on the longitudinal polarisability is very narrow, while the NLO band on the Baldin sum rule is significantly wider. In the latter case, one can, again, argue that the effects of higher orders might be overestimated towards $Q=0$ (even though the band is barely visible there in this scale), since we know that the dominant $\alpha_{E1}$ contribution is well reproduced already at NLO (cf.\ also the discussion regarding $\beta_{M1}$ above); however, this is not any longer the case at larger finite virtualities, where more sizeable contributions could be expected, especially from higher-order magnetic couplings.

\begin{figure}[htb]
    \centering
    \includegraphics[width=0.491\textwidth]{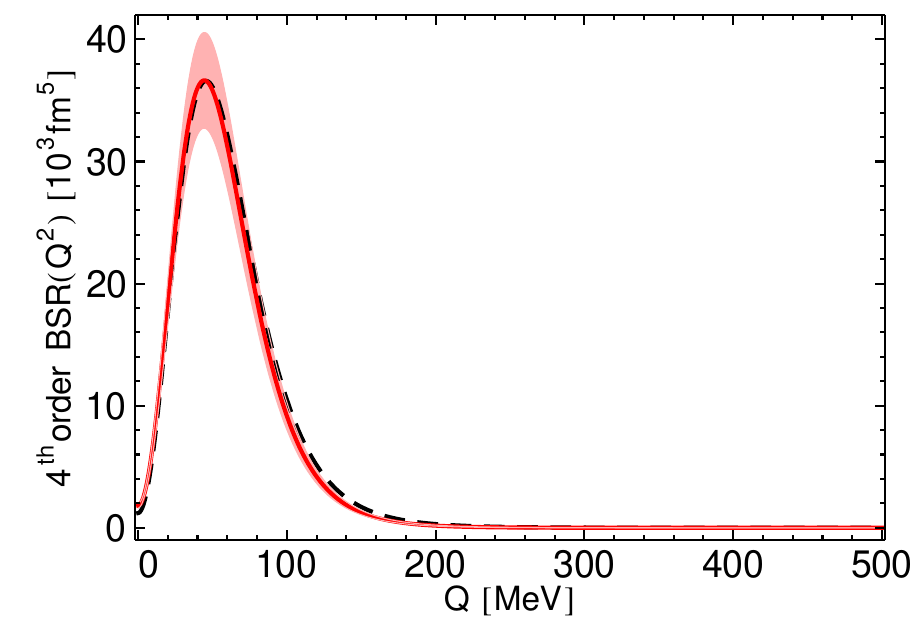}
    \caption{Fourth-order generalised Baldin sum rule. The curves are coded as in the right panel of Fig.~\ref{fig:alpha_beta}.}
    \label{fig:BSR4}
\end{figure}

Finally, we consider the fourth-order generalised Baldin sum rule, defined in terms of the transverse amplitude as~\cite{Lensky:2017bwi}
\begin{align}
\left[\alpha_{E1,\nu}+\beta_{M1,\nu}+\nicefrac{1}{12}(\alpha_{E2}+\beta_{M2})\right](Q^2)=\frac{1}{8\pi}\frac{\mathrm{d}^2f_T(\nu,Q^2)}{\mathrm{d}(\nu^2)^2}\bigg|_{\nu=0}\,.
\end{align}
The static limit value of that sum rule gives the corresponding linear combination of the dispersive ($\alpha_{E1,\nu}$, $\beta_{M1,\nu}$) and quadrupole ($\alpha_{E2}$, $\beta_{M2}$) electric and magnetic polarisabilities of the deuteron, defined as in Ref.~\cite{Babusci:1998ww,Holstein:1999uu}.\footnote{The quadrupole polarisabilities (spin independent) are not to be confused with the tensor polarisabilities of the deuteron defined in, e.g., Refs.~\cite{Chen:1998vi,Ji:2003ia}.} Analogously to the Baldin sum rule considered above, using the exact formula $|\bv{q}|=\sqrt{\nu^2+Q^2}$ should allow one to recover the subleading polarisabilities here as well. Expanding the value at $Q=0$ up to N3LO, we get, order-by-order:
\begin{align}
\alpha_{E1,\nu}+\beta_{M1,\nu}+\nicefrac{1}{12}(\alpha_{E2}+\beta_{M2})
&=\left[0.865 + 0.597 + 0.345 + 0.007\right]\times 10^3\text{ fm}^5\nonumber\\
&=1.814\times 10^3\text{ fm}^5\,.
\end{align}
We refrain from showing here the analytic expression due to its length. The plot of the fourth-order generalised Baldin sum rule is shown in Fig.~\ref{fig:BSR4}. One can see that it shares the features with the Baldin sum rule, sharply rising from the static limit and peaking at finite values of $Q$, and also getting a rather small NLO contribution at finite virtualities. It has to be noted that such a rapid growth of the subleading coefficients in the expansion of $f_T(\nu,Q^2)$ in powers of $\nu^2$, i.e., the Baldin sum rule and its fourth-order analogue, at finite $Q$ can be of some concern in the context of a calculation of the TPE correction in $\mu$D, and it is important to verify that the transverse contribution is small (according to what is given by the counting, and is indeed seen in dispersive calculations~\cite{Acharya:2020bxf,Hernandez:2019zcm}).

\section{Conclusion}
\label{sec:Conclusion}
We have calculated the unpolarised deuteron VVCS amplitudes in the framework of \piEFT/, using the $z$-parametrisation. Our results are at N3LO in the \piEFT/ expansion for the longitudinal amplitude, and at NLO for the transverse amplitude.  We have provided analytic expressions for the VVCS amplitudes --- the possibility to do so being one of the advantages of the \piEFT/ framework.

Investigating the RG running of the longitudinal amplitude, we show that there is a single unknown one-photon two-nucleon contact term contributing to the amplitude at N3LO, parametrised by the low-energy constant $l_1^{C0_S}$. This constant is extracted from a fit to the deuteron charge form factor also at N3LO; the corresponding \piEFT/ result is obtained from the residue of the longitudinal amplitude.
We note that two of the RG equations for the N3LO two-nucleon one-photon contact terms obtained by us are at variance with those obtained in Ref.~\cite{Chen:1999bg}. Since these RG equations simply relegate these contact terms to higher orders, this disagreement alone should not make the results of Ref.~\cite{Chen:1999bg} incompatible with ours. On the other hand, the calculation of Ref.~\cite{Rupak:1999rk} that deals with the same contact terms at N4LO could be affected. However, a re-evaluation of its results might still not be needed. Namely, we suspect the reason for the difference in the RG equations to be a factor of 2 missing in  Refs.~\cite{Chen:1999bg,Rupak:1999rk} in all contributions of the contact terms in question. A more serious mistake would likely leave RG scale dependent results, which do not seem to appear. Fitting an RG-invariant quantity to an experimental datum should renormalise the missing constant factor, leading to the correct result for the observables.

The deuteron charge and magnetic form factors, calculated, respectively, at N3LO and NLO from the residues of the longitudinal and transverse VVCS amplitudes, recover the previously obtained NNLO result for the charge form factor~\cite{Phillips:1999hh}, as well as the NLO result for the magnetic form factor~\cite{Chen:1999tn}.
We furthermore obtain the generalised deuteron polarisabilities, namely, the electric $\alpha_{E1}(Q^2)$ and magnetic $\beta_{M1}(Q^2)$ dipole polarisabilities, the longitudinal polarisability $\alpha_L(Q^2)$, and the generalised Baldin sum rule $\left[\alpha_{E1}+\beta_{M1}\right](Q^2)$. We also calculate the fourth-order generalised Baldin sum rule.
The results for the electric and longitudinal generalised polarisability are at N3LO, while the remaining ones are calculated at NLO.
There is a slight discrepancy between our expression for the static electric polarisability of the deuteron $\alpha_{E1}$ and the results obtained previously in Ref.~\cite{Phillips:1999hh}. The numerical values of the static electric and magnetic polarisabilities of the deuteron, $\alpha_{E1}$ and $\beta_{M1}$, are in a good agreement with other calculations, in particular, the recent $\chi$EFT-based evaluation in Ref.~\cite{Acharya:2020bxf}. This agreement between the values of $\alpha_{E1}$ calculated in the two low-energy effective field theories would be expected, since $\alpha_{E1}$ is dominated by low-energy (or long-range) properties of the deuteron. On the other hand, the fact that the results for $\beta_{M1}$ agree so well between the two theories can be reasonably explained by the properties of the procedure used to determine the NLO isovector magnetic contact term in Ref.~\cite{Rupak:1999rk}, namely, that it reproduced the cross section in $np\to d\gamma$ at low energies (although this remains to be explicitly demonstrated). The results of this work, in particular, the longitudinal amplitude, can be used to benchmark other calculational frameworks, e.g., $\chi$EFT potentials, especially in settings where reproducing low-energy properties of the deuteron is important, such as the calculation of the TPE corrections in $\mu$D.

The N3LO \piEFT/ result for the longitudinal amplitude and the charge form factor provides high-precision model-independent input for a calculation of the TPE correction to the Lamb shift in $\mu$D. The uncertainty of such a calculation will be dominated by higher-order terms in the \piEFT/ expansion, expected to be of the order of $1\%$ at N3LO. At the same time, the result for $f_T(\nu,Q^2)$ can be used to verify the smallness of the transverse contribution to the TPE correction. One has to point out that the numerically most important corrections are likely to come from high-order terms in the one-nucleon sector, such as those stemming from the expansion of the nucleon elastic form factors and the single-nucleon VVCS amplitudes, owing to the large energy and momentum scales needed in order to probe the nucleon structure and the resulting slow convergence of the said expansion. The investigation of these issues, together with a detailed analysis of the uncertainty of the TPE correction, will be presented in a subsequent publication~\cite{muDpaper}. We are also investigating the spin-dependent deuteron VVCS amplitudes and the corresponding generalised polarisabilities of the deuteron.

A natural extension of this work would involve an application to unpolarised VVCS in $A=3$ nuclei, where a good description of quantities such as the charge radii has already been obtained in \piEFT/, see, e.g., Refs.~\cite{Vanasse:2015fph,Vanasse:2017kgh} for recent results, and Ref.~\cite{Hammer:2019poc} for a review. Given the complications of \piEFT/ as applied to the three-nucleon sector, pursuing this program would not be straightforward, however, it promises interesting results.

\section*{Acknowledgements}
We thank V.~Baru and A.~Filin for discussing the details of their work and for sharing with us the results of their $\chi$EFT calculation of the deuteron charge form factor. We thank C.~Carlson and M.~Gorchtein for useful communications. The calculations in this work were performed with the help of \textsc{FORM}~\cite{Vermaseren:2000nd}, and the figures in the article were made with the help of \textsc{JaxoDraw}~\cite{Binosi:2003yf} and \textsc{SciDraw}~\cite{Caprio:2005dm}.

This work was supported by the Deutsche Forschungsgemeinschaft (DFG, German Research Foundation), 
in part through the Collaborative Research Center [The Low-Energy Frontier of the Standard Model, Projektnummer 204404729 - SFB 1044], 
and in part through the 
Cluster of Excellence [Precision Physics, Fundamental Interactions, and Structure of Matter] (PRISMA$^+$ EXC 2118/1) 
within the German Excellence Strategy (Project ID 39083149). It was also supported by the U.S. Department of Energy contract DE-AC05-06OR23177, under which Jefferson Science Associates, LLC, manages and operates Jefferson Lab.

\end{document}